\documentclass[journal=jctcce,manuscript=article]{achemso}
\setkeys{acs}{articletitle = true}
\setkeys{acs}{chaptertitle = true}
\setkeys{acs}{abbreviations = false}
\setkeys{acs}{maxauthors=150}
\setkeys{acs}{etalmode=truncate}

\usepackage{graphicx}
\usepackage{multirow}
\usepackage{siunitx}
\usepackage{xcolor}
\usepackage{braket}
\usepackage{bm}
\usepackage{amsmath}
\usepackage{float}

\usepackage[hidelinks]{hyperref}
\hypersetup{colorlinks=false}

\usepackage{numprint}
\usepackage[utf8]{inputenc}
\usepackage[T1]{fontenc}
\usepackage{mathptmx}
\usepackage[english]{babel}
\usepackage{subfigure}
\usepackage{comment}
\usepackage{siunitx}

\usepackage{xr}
\DeclareMathAlphabet{\mathcal}{OMS}{cmsy}{m}{n}
\title{Reduced density matrix formulation of quantum linear response} 

\author{Theo Juncker von Buchwald}
\email{tjvbu@kemi.dtu.dk}
\affiliation{Department of Chemistry, Technical University of Denmark, Kemitorvet Building 207, DK-2800 Kongens Lyngby, Denmark.}
\author{Karl Michael Ziems}
\email{kmizi@kemi.dtu.dk}
\affiliation{Department of Chemistry, Technical University of Denmark, Kemitorvet Building 207, DK-2800 Kongens Lyngby, Denmark.}
\author{Erik Rosendahl Kjellgren}
\email{kjellgren@sdu.dk}
\affiliation{Department of Physics, Chemistry and Pharmacy,
University of Southern Denmark, Campusvej 55, 5230 Odense, Denmark.}
\author{Stephan P. A. Sauer}
\affiliation{Department of Chemistry, University of Copenhagen, DK-2100 Copenhagen \O.}
\author{Jacob Kongsted}
\affiliation{Department of Physics, Chemistry and Pharmacy,
 University of Southern Denmark, Campusvej 55, 5230 Odense, Denmark.}
\author{Sonia Coriani}
\affiliation{Department of Chemistry, Technical University of Denmark, Kemitorvet Building 207, DK-2800 Kongens Lyngby, Denmark.}

\date{April 25, 2024}

\begin{document}

\begin{abstract}

The prediction of spectral properties via linear response (LR) theory is an important tool in quantum chemistry for understanding photo-induced processes in molecular systems. With the advances of quantum computing, we recently adapted this method for near-term quantum hardware using a truncated active space approximation with orbital rotation, named quantum linear response (qLR). In an effort to reduce the classic cost of this hybrid approach, we here derive and implement a reduced density matrix (RDM) driven approach of qLR. This allows for the calculation of spectral properties of moderately sized molecules with much larger basis sets than so far possible. We report qLR results for benzene and $R$-methyloxirane with a cc-pVTZ basis set and study the effect of shot noise on the valence and oxygen K-edge absorption spectra of H$_2$O in the cc-pVTZ basis.

\end{abstract}

\maketitle

\section{\label{sec:intro}Introduction}

The ability to predict molecular properties such as excitation energies, oscillator and rotational strengths, is important for the interpretation of spectroscopic data, and a cornerstone in many scientific contexts, e.g., 
photochemistry, photophysics,
photocatalysis, and 
optogenetics. A well-established framework to calculate such properties (on conventional hardware) is molecular response theory \cite{olsen1985linear,Christiansen1998,helgaker2012recent,Pawlowski2015}.
Within such theory, the response of the molecular system 
to (frequency-dependent) perturbing fields 
is rationalized in terms of linear and non-linear 
response functions~\cite{olsen1985linear,Christiansen1998,helgaker2012recent,Pawlowski2015}. Through the 
calculation of the linear response function, for example, one can obtain the electric dipole
polarizability and the optical rotation tensor. From the poles of the
frequency-dependent response functions
one can get the excited state energies, whereas
the residues yield various transition properties (e.g. for multi-photon absorption processes) and properties of excited states.
Response theory has been implemented for most electronic structure methods, including Hartree-Fock self-consistent field (SCF)~\cite{olsen1985linear,Norman1995}, multi-configurational SCF (MCSCF)~\cite{olsen1985linear,Hettema1992,Jonsson1996}, coupled cluster
(CC)~\cite{koch1990,Christiansen1998}, Møller-Plesset perturbation theory \cite{jod05-jpca109-11618} and time-dependent density functional theory~\cite{Salek2002,Parker2018}.

With the ongoing promising developments in quantum computing and its expected impact on quantum chemistry, it comes as no surprise that linear response theory 
has recently been expanded to quantum hardware, 
and named 
quantum linear response (qLR) \cite{Kumar2023, ziems2023options}. In the approach by \citeauthor{Kumar2023} \cite{Kumar2023}, the authors perform wave function optimization and linear response in the complete orbital space to the level of singles and doubles. Therein, they introduce two full-space qLR approaches, namely self-consistent qLR and projected qLR. However, since current quantum hardware (referred to as noisy intermediate-scale quantum, NISQ) is extremely prone to errors due to noise and decoherence, chemical calculations are restricted to only a few qubits and 
necessitate hybrid approaches \cite{chen2023complexity,mcclean2016theory}. The latter are defined as splitting the algorithm workload into quantum and classic parts, outsourcing less correlated calculations to classical machines.  

The qLR approach by 
\citeauthor{ziems2023options}~\cite{ziems2023options} builds upon this premise 
with a NISQ-era formulation of qLR. 
Here, the authors combine an active space approximation with truncated excitation ranks and orbital optimization. The hybrid approach comes into play by only mapping the active space contributions onto the quantum computer. 
By doing so, they can dramatically reduce the number of parameters needed to describe the linear response properties and, importantly, can rely on less converged wave functions, which translates into shallower circuits and allows for the treatment of larger molecules and basis sets on near-term hardware. In their work,  
\citeauthor{ziems2023options}
introduce various parametrization schemes for the orbital rotation and active space excitation operators. This yields eight different qLR methods, three of which are classified as near-term. These are the naive, proj, and all-proj qLR ans{\"a}tze with active spaces. Note that this qLR approach is similar to the complete active space self-consistent field (CASSCF) method \cite{yeager1979multiconfigurational,dalgaard1980time,roos1980complete,Siegbahn1980,Siegbahn1981} commonly used on classical hardware with the difference of truncating the active space excitation rank and the use of various novel parameterization schemes. Closely related to the qLR approaches are also the qEOM approaches \cite{ollitrault_quantum_2020, jensen_qeom_2024}.

Our goal here is to reformulate the qLR approach with active spaces by \citeauthor{ziems2023options}~\cite{ziems2023options} using reduced density matrices (RDMs). RDMs have been used in the past to efficiently formulate approaches on classical hardware resulting in reduced computational scaling \cite{fosso2016large,mazziotti2008parametrization,mcweeny1956density,rice1985efficient}. This is especially important for the classical part of the hybrid approach behind qLR. With increasing basis set size, the orbital rotation parameters can become the bottleneck over the quantum part. Using RDMs will allow performing qLR with even larger basis sets and reduced classical costs. 

When truncating the linear response active space excitation rank above singles (qLRS), the naive qLRSD is the only ansatz that requires no more than 
4-electron RDMs. The other near-term qLR approaches, proj qLRSD and all-proj qLRSD, both require 5- and 6-electron RDMs, which becomes quickly unfeasible due to the exponential memory scaling of the RDMs. On this basis, in this work, the naive qLRSD 
ansatz with RDMs is derived and subsequently compared against classic CASSCF LR for the absorption spectrum of benzene, using an active space of 6 electrons in 6 active orbitals [in short (6,6)], and the electronic circular dichroism (ECD) spectrum of $R$-methyloxirane, also with a (6,6) active space. Using the RDM ansatz of naive qLRSD allows us to simulate these systems up to and including the cc-pVTZ basis set. Additionally, a shot noise analysis investigation of the RDM qLR formulation is performed on H$_2$O (4,4)/cc-pVTZ.\\

This paper is organized as follows.
In Section~\ref{sec:theo}, we 
shortly review 
unitary coupled cluster theory as well as the active space approximation and linear response. We then introduce reduced density matrices and briefly discuss the scaling of our RDM-driven naive qLR approach. 
In Section~\ref{sec:comp}, we provide computational details, followed by Section~\ref{sec:results} where we present 
and discuss the results for the absorption spectrum of benzene (Section \ref{ssec:benzene}) and the 
ECD spectrum of $R$-methyloxirane (Section \ref{ssec:methyloxirane}) using the RDM formalism of linear response and compare these against CASSCF results. 
In Section~\ref{ssec:shot_noise}, we investigate the effects of shot noise when using the RDM-based naive qLR approach. 
Finally,  some concluding remarks are given in Section \ref{sec:summary}.

\section{\label{sec:theo}Theory}

In this section we 
formulate and provide 
working equations for a reduced density matrix-based approach to linear response theory with a truncated active space. We start by introducing the orbital-optimized unitary coupled cluster (oo-UCC) method in 
Sec.~\ref{ssec:UCC}, followed by an introduction to the active space decomposition in Sec.~\ref{ssec:active_space}, and the linear response theory in an active space framework in Sec. \ref{ssec:LR}. After this, a brief overview of reduced density matrices (RDMs) is given in Sec.~\ref{ssec:RDM} and a short discussion of scaling in Sec. \ref{ssec:scaling}. It is important to note that the linear response equations derived here may be utilized with any variational reference state and not just oo-UCC. Throughout the manuscript, unless otherwise stated, this index notation will be used: \textit{p, q, r, s, t, u, m,} and \textit{n} are general indices; \textit{a, b, c,} and \textit{d} are virtual indices; \textit{i, j, k,} and \textit{l} are inactive indices; and $\nu_a$ and $\nu_i$ are active space orbitals that are, respectively, virtual and inactive in the Hartree-Fock reference state.

\subsection{Unitary coupled cluster}
\label{ssec:UCC}

In unitary coupled cluster (UCC) theory \cite{bartlett1989alternative}, the UCC wave function is given by an exponential ansatz of a reference state
\begin{equation}
    \ket{\text{UCC}} = \text{e}^{\hat T(\theta) - \hat T^\dagger(\theta)} \ket{0}
\end{equation}
where $\hat T(\theta)$ is the cluster operator and $\hat T^\dagger(\theta)$ is its Hermitian conjugate, both depending on the quantum circuit parameters $\theta$. The cluster operator can be expanded by excitation ranks
\begin{equation}
    \hat T(\theta) = \hat T_1(\theta) + \hat T_2(\theta) + \ldots
\end{equation}
where
\begin{align}
    \hat T_1(\theta) = & \sum_{ai}\theta_{ai} \hat E_{ai} \\\nonumber
    \hat T_2(\theta) = & \frac{1}{2} \sum_{aibj}\theta_{aibj} \hat E_{ai} \hat E_{bj} \\\nonumber
    & \vdots
\end{align}
and $\hat{E}_{ai}$ is the singlet one-electron excitation operator. The UCC energy expression is
\begin{equation}
    E(\theta) = \braket{\text{UCC}(\theta) | \hat H | \text{UCC}(\theta)}
\end{equation}
with the Hamiltonian defined as
\begin{equation}
    \hat{H} = \sum_{pq}h_{pq}\hat{E}_{pq} + \frac{1}{2}\sum_{pqrs}g_{pqrs}\hat{e}_{pqrs}
\end{equation}
where $\hat e_{pqrs}$ is the two-electron excitation operator, defined as
\begin{align}
    \hat e_{pqrs} = & \hat E_{pq} \hat E_{rs} - \delta_{rq} \hat E_{ps}~;
\end{align}
$h_{pq}$ and $g_{pqrs}$ are the one- and two-electron integrals in the molecular orbital (MO) basis. 
The $\theta$ circuit parameters are found by minimizing the energy with respect to the circuit parameter values.
A continuation of the UCC formalism is the orbital-optimized UCC (oo-UCC) \cite{Mizukami2020,Sokolov2020}, where the UCC wavefunction is exponentially parameterized by the orbital rotation 
operator
\begin{equation}
\label{kappa}
    \hat{\kappa}\left(\boldsymbol{\kappa}\right) = \sum_{p>q}\kappa_{pq}\hat{E}^-_{pq}~,
\end{equation}
with
\begin{equation}
    \hat E^-_{pq} = \hat E_{pq} - \hat E_{qp}~,
\end{equation}
as
\begin{equation}
    \ket{\text{oo-UCC}(\kappa,\theta)} = \text{e}^{-\hat{\kappa}\left(\boldsymbol{\kappa}\right)} \ket{\text{UCC}(\theta)}~.
\end{equation}

When combined with UCC in an active space, the only non-redundant rotations are \textit{inactive to active} $(AI)$, \textit{inactive to virtual} $(VI)$, and \textit{active to virtual} $(VA)$, $pq\in\{AI,VI,VA\}$. Here, $I$, $A$, and $V$ refer to the set of orbitals contained in, respectively, the inactive, active, and virtual space.
Instead of including the orbital rotation in the wave function, it is convenient to transform the Hamiltonian (integrals) using the orbital rotation parameters
\begin{equation}
    \hat{H}\left(\kappa\right) = \sum_{pq}h_{pq}\left(\kappa\right)\hat{E}_{pq} + \frac{1}{2}\sum_{pqrs}g_{pqrs}\left(\kappa\right)\hat{e}_{pqrs}
\end{equation}
using
\begin{align}
    h_{pq}\left(\kappa\right) &= \sum_{p'q'} \left[\text{e}^\kappa\right]_{q'q}h_{p'q'}\left[\text{e}^{-\kappa}\right]_{p'p}\\
    g_{pqrs}\left(\kappa\right) &= \sum_{p'q'r's'}\left[\text{e}^\kappa\right]_{s's}\left[\text{e}^\kappa\right]_{q'q}g_{p'q'r's'}\left[\text{e}^{-\kappa}\right]_{p'p}\left[\text{e}^{-\kappa}\right]_{r'r}.
\end{align}
The oo-UCC energy expression is then
\begin{equation}
    E(\theta) = \braket{\text{UCC}(\theta) | \hat H(\kappa) | \text{UCC}(\theta)}.
    \label{eq:oo-UCC_energy}
\end{equation}
Performing a minimization of the energy with respect to the circuit parameters, which are optimized on the quantum hardware, and the orbital rotation parameters, which are split into classic and quantum workload, is known as the orbital optimized variational quantum eigensolver (oo-VQE) \cite{Mizukami2020,Sokolov2020}.

\subsection{Active space}
\label{ssec:active_space}

A well-known 
active space method in `classic' quantum chemistry is CASSCF \cite{Siegbahn1980,roos1980complete,Siegbahn1981}. In CASSCF, the full space is split into subspaces consisting of an inactive space where all orbitals are doubly occupied, an active space that is treated using a complete configuration interaction (CI) expansion, and a virtual space where all orbitals are empty. In the quantum approach we use here, the CI expansion in the active space is truncated to singles and doubles excitations \cite{ziems2023options}. Applying an active space approximation to the wave function yields
\begin{equation}
        \left|0\left(\theta\right)\right> = \left|I\right>\otimes \left|A\left(\theta\right)\right> \otimes \left|V\right>~,
\end{equation}
where $\ket{I}$ is the inactive part of the wave function, $\ket{A(\theta)}$ is the active part, and $\ket{V}$ is the virtual part. The active space is parameterized by $\theta$ by using a unitary transformation
\begin{equation}
    \ket{A(\theta)} = U(\theta)\ket{A}
\end{equation}
Operators can likewise be decomposed as
\begin{equation}
    \hat{O} = \hat{O}_I\otimes \hat{O}_A\otimes \hat{O}_V.
\end{equation}
By applying the active space approximation, expectation values of any operator now have the form
\begin{equation}
    \braket{0(\theta) | \hat O | 0(\theta)} = \braket{I | \hat O_I | I} \braket{A(\theta) | \hat O_A | A(\theta)} \braket{V | \hat O_V | V}. \label{eq:AS_seperation}
\end{equation}
Using the active space approximation on the oo-UCC energy expression [Eq. \eqref{eq:oo-UCC_energy}], only the active space part is calculated on the quantum hardware. 

\subsection{Linear response}
\label{ssec:LR}

Here, we adopt the qLR approach with active spaces introduced by \citeauthor{ziems2023options}~\cite{ziems2023options}
that uses, as a starting point, the MCSCF wave function, 
\begin{equation}
    \ket{\tilde{0}(t)} = \text{e}^{\hat\kappa(t)}\text{e}^{\hat S(t)}\ket{0}
\end{equation}
\begin{align}
    \hat\kappa(t) & = \sum_\mu \left( \kappa_\mu(t) \hat q_\mu + \kappa_\mu^*(t)\hat q_\mu^\dagger \right) \label{eq:kappa_t} \\
    \hat S(t) & = \sum_n \left( S_n(t) \hat G_n + S_n^*(t)\hat G_n^\dagger \right) \label{eq:s_t}
\end{align}
where the wave function $\ket{0}$ is parameterized through the time-dependent orbital operator $\hat{\kappa}(t)$ and active space rotation operator $\hat{S}(t)$, which contain, respectively, the orbital rotation operators 
$\hat q_\mu$ and the active space excitation operators $\hat G_n$. \citeauthor{ziems2023options} then introduce various ans{\"a}tze for linear response on near-term quantum computers using a variety of choices for $\hat q_\mu$ and $\hat G_n$ \cite{ziems2023options}. Here, we specifically focus on the naive operators
\begin{align}
    \hat q_{pq} = \frac{1}{\sqrt{2}}\hat{E}_{pq} \label{eq:Q}
\end{align}
\begin{align}
    \hat{G} \in \Bigg\{\frac{1}{\sqrt{2}}\hat{E}_{v_av_i},\quad &\frac{1}{2\sqrt{\left(1+\delta_{v_av_b}\right)\left(1+\delta_{v_iv_j}\right)}}\left(\hat{E}_{v_av_i}\hat{E}_{v_bv_j} + \hat{E}_{v_av_j}\hat{E}_{v_bv_i}\right), \label{eq:R} \\\nonumber
    &\frac{1}{2\sqrt{3}}\left(\hat{E}_{v_av_i}\hat{E}_{v_bv_j} - \hat{E}_{v_av_j}\hat{E}_{v_bv_i}\right)\Bigg\}~.
\end{align}
We will refer to the method as naive qLRSD, since only singles and doubles excitations are included in the linear response excitation space.

The key quantity in linear response theory is the frequency-dependent linear response function 
$\braket{\braket{\hat A;\hat B}}_{\omega_B}$, 
for two generic operators $\hat{A}$ and $\hat{B}$.
The linear response function
can be computed as~\cite{helgaker2012recent}
\begin{align} \label{eq:LR_equation}
\braket{\braket{\hat A;\hat B}}_{\omega_B} 
&= 
-\textbf{V}_A^{[1]\dagger}
\boldsymbol{\beta}_{B}.
\end{align}
This requires solving sets of linear response equations~\cite{olsen1985linear,helgaker2012recent,Jorgensen1988} 
\begin{equation}
    \left( \textbf{E}^{[2]} - \omega_B \textbf{S}^{[2]} \right) \boldsymbol{\beta}_B = \textbf{V}^{[1]}_B \label{eq:LR_eq}
\end{equation}
where $\textbf{E}^{[2]}$ is the electronic Hessian, $\textbf{S}^{[2]}$ is the Hermitian metric, $\boldsymbol{\beta}_B$ is the linear response vector, and $\textbf{V}^{[1]}_B$ is the property gradient. The Hessian and Hermitian metric can be expressed as
\begin{align}
 \textbf{E}^{[2]} &= \begin{pmatrix}
    \boldsymbol{A} & \boldsymbol{B} \\
      \boldsymbol{B}^* & \boldsymbol{A}^*           
     \end{pmatrix}, \quad 
     \textbf{S}^{[2]} = \begin{pmatrix}
    \boldsymbol{\Sigma} & \boldsymbol{\Delta} \\
     -\boldsymbol{\Delta} ^* &  -\boldsymbol{\Sigma}^*           
     \end{pmatrix} \quad
\end{align}
with the submatrices
\begin{align}
    \boldsymbol{A} &= \begin{pmatrix} \label{eq:LR_A}
\left<0\left|\left[\hat{q}_\mu^\dagger,\left[\hat{H},\hat{q}_{\nu}\right]\right]\right|0\right>
& \left<0\left|\left[\hat{G}_{n},\left[\hat{H},\hat{q}^\dagger_\nu\right]\right]\right|0\right> \\
\left<0\left|\left[\hat{G}_{n}^\dagger,\left[\hat{H},\hat{q}_{\nu}\right]\right]\right|0\right>
& \left<0\left|\left[\hat{G}_{n}^\dagger,\left[\hat{H},\hat{G}_{m}\right]\right]\right|0\right>
\end{pmatrix} \\
    \boldsymbol{B} &= \begin{pmatrix} \label{eq:LR_B}
\left<0\left|\left[\hat{q}_\mu^\dagger,\left[\hat{H},\hat{q}_{\nu}^\dagger\right]\right]\right|0\right>
& \left<0\left|\left[\hat{G}^\dagger_{n},\left[\hat{H},\hat{q}^\dagger_\nu\right]\right]\right|0\right> \\
\left<0\left|\left[\hat{G}_{n}^\dagger,\left[\hat{H},\hat{q}_{\nu}^\dagger\right]\right]\right|0\right>
& \left<0\left|\left[\hat{G}_{n}^\dagger,\left[\hat{H},\hat{G}_{m}^\dagger\right]\right]\right|0\right>
\end{pmatrix} \\
    \boldsymbol{\Sigma} &= \begin{pmatrix} \label{eq:LR_sigma}
\left<0\left|\left[\hat{q}_\mu^\dagger,\hat{q}_{\nu}\right]\right|0\right>
& \left<0\left|\left[\hat{q}_{\mu}^\dagger,\hat{G}_{m}\right]\right|0\right> \\
\left<0\left|\left[\hat{G}_{n}^\dagger,\hat{q}_{\nu}\right]\right|0\right>
& \left<0\left|\left[\hat{G}_{n}^\dagger,\hat{G}_{m}\right]\right|0\right>
\end{pmatrix} \\
    \boldsymbol{\Delta} &= \begin{pmatrix} \label{eq:LR_delta}
\left<0\left|\left[\hat{q}_\mu^\dagger,\hat{q}_{\nu}^\dagger\right]\right|0\right>
& \left<0\left|\left[\hat{q}_{\mu}^\dagger,\hat{G}_{m}^\dagger\right]\right|0\right> \\
\left<0\left|\left[\hat{G}_{n}^\dagger,\hat{q}_{\nu}^\dagger\right]\right|0\right>
& \left<0\left|\left[\hat{G}_{n}^\dagger,\hat{G}_{m}^\dagger\right]\right|0\right>
\end{pmatrix}.
\end{align}
The property gradient has the form
\begin{align}
     \textbf{V}^{[1]}_B = \begin{pmatrix}
     \braket{0|[\hat{q}_\mu,\hat{B}]|0} \\[0.3em]
     \braket{0|[\hat{G}_n,\hat{B}]|0} \\[0.3em]
     \braket{0|[\hat{q}^\dagger_\mu,\hat{B}]|0} \\[0.3em]
     \braket{0|[\hat{G}^\dagger_n,\hat{B}]|0} \\[0.3em]
     \end{pmatrix}.
\end{align}
The linear response vector is given as
\begin{align}
    \boldsymbol{\beta}_B = \begin{pmatrix}
    \boldsymbol{Z}_B \\
      \boldsymbol{Y}_B^*            
     \end{pmatrix} = \begin{pmatrix}
    \boldsymbol{\kappa}_B \\
      \boldsymbol{S}_B \\
      \boldsymbol{\kappa}_{-B}^* \\
      \boldsymbol{S}_{-B}^* \\
     \end{pmatrix}
\end{align}
where $\kappa_B$ and $S_B$ are the first order expansions of the
orbital parameters 
in Eqs. \eqref{eq:kappa_t} and \eqref{eq:s_t}. 
The spectral representation of the linear response function can be written as
\begin{align}
\langle\langle 
\hat{A};
\hat{B}
\rangle\rangle_{\omega_B}
&=
\sum_{k>0} \frac{\braket{0| [\hat{A},\hat{\tilde{O}}_k] |0}\braket{0 | [\hat{\tilde{O}}_k^\dagger,\hat{B}] |0}}{\omega-\omega_k} - \sum_{k>0} \frac{\braket{0| [\hat{B},\hat{\tilde{O}}_k] |0}\braket{0 | [\hat{\tilde{O}}_k^\dagger,\hat{A}] |0}}{\omega+\omega_k}~,\label{eq:pol}
\end{align}
where $\hat{\tilde O}_k$ is a normalized (excitation) 
operator
\begin{align}
    \hat{\tilde{O}}_k = & \frac{\hat{O}_k}{\sqrt{\braket{k|k}}} = \frac{\hat{O}_k}{\sqrt{\braket{ 0|[ \hat{O}_k, \hat{O}_k^\dagger ] |0}}} \\
    \hat{O}_k = & \sum_{l \in \mu,n}\left({Z}_{k,l} \hat{X}_l^\dagger + Y_{k,l}\hat{X}_l\right). \label{eq:exc_vec}
\end{align}
In the equation above, $\hat X_l$ is the set of orbital rotation and active space excitation operators $\{ \hat q_\mu, \hat G_n \}$,
and $Z_{k,l}$ and $Y_{k,l}$ are 
their weights.
The (one-electron) operators 
$\hat{A}$ and $\hat{B}$ are in 
their second quantization form, e.g.
\begin{equation}
    \hat{A} = 
    \sum_{pq} 
    {\mathcal{A}}_{pq}
    \hat{E}_{pq}~, 
    \end{equation}
where ${\mathcal{A}}_{pq}$ are the $\hat{A}$ operator's integrals over the MO basis.

From the poles and residues of the linear response function, transition properties like excitation energies and transition strengths are obtained. The excitation energies and excitation vectors are found by solving the eigenvalue equation
\begin{equation}
    \textbf{E}^{[2]}\boldsymbol{\beta}_k = \omega_k \textbf{S}^{[2]}\boldsymbol{\beta}_k. \label{eq:LR_exc}
\end{equation}
which is the homogeneous version of the linear response equations in Eq.~\eqref{eq:LR_equation}, and where $\boldsymbol{\beta}_k$ is the excitation vector containing the $Z_{k,l}$ and $Y_{k,l}$ parameters in Eq. \eqref{eq:exc_vec}.
The transition strengths are the residues of the linear response function and are found according to
\begin{align}
    \lim_{\omega \rightarrow \omega_k} (\omega-\omega_k)\braket{\braket{\hat{A} ; \hat{B}}}_{\omega} = \braket{0| [\hat{A},\hat{\tilde{O}}_k] |0}\braket{0 | [\hat{\tilde{O}}_k^\dagger,\hat{B}] |0} = \left(\textbf{V}^{[1]\dagger}_A  \boldsymbol{\beta}_k\right) \left(\boldsymbol{\beta}_k^\dagger\textbf{V}^{[1]}_B \right)
\end{align}
which, as seen, requires the property gradient and the excitation vector. 

A prototypical example of a linear response function is the dipole polarizability
\begin{align}
\alpha_{\gamma\delta}(\omega) &= 
- \langle\langle 
\hat{\mu}_\gamma;
\hat{\mu}_\delta
\rangle\rangle_\omega
\end{align}
where $\hat \mu_\gamma$ is a Cartesian component of the one-electron electric dipole moment operator ($\gamma \in \{x,y,z\}$),  
in second quantization given by
\begin{align}
    \hat \mu_\gamma = 
    - \sum_{pq} 
    (\vec{r}_\gamma)_{pq}
    \hat{E}_{pq} \\
    (\vec{r}_\gamma)_{pq} = \left\langle p\left| \vec{r}_\gamma \right|q \right\rangle.
\end{align} 
\noindent In this paper, we focus on the oscillator strengths $f_k$ of one-photon absorption, which are related to the residues of the electric dipole polarizability,
\begin{align}
    f_k &=  \frac{2}{3} \omega_k \sum_\gamma \big| \braket{0| [\hat{\mu}_\gamma,\hat{\tilde{O}}_k] |0} \big|^2~,
\end{align}
and on the rotational strengths $R_k$ of ECD, connected to the residues of the optical rotation tensor, 
\begin{align}
R_k = {\text{Im}}
 (
 \langle 0 | [\hat{
 \boldsymbol{\mu}},\hat{\tilde{O}}_k] | 0 \rangle 
 \cdot
\braket{0 |
[\hat{
\tilde{O}}_k^\dagger,\hat{\bm{m}}] | 0}
)
\end{align}
where $\hat{\bm{m}}$ is the magnetic dipole moment operator
\begin{align}
    \hat{\bm{m}} = - \frac{i}{2} \sum_{pq} 
    (\vec r \times \vec \nabla)_{pq}
    \hat{E}_{pq} \\
    (\vec r \times \vec \nabla)_{pq} = \left\langle p\left| \vec r \times \vec \nabla \right|q \right\rangle.
\end{align}

\subsection{Reduced density matrices}
\label{ssec:RDM}

As anticipated, this work formulates naive qLRSD in terms of reduced density matrices. This entails calculating the Hessian, the metric and the property gradients of the linear response equation [Eq. \eqref{eq:LR_eq}] using RDMs. To this end, the largest terms we can obtain from Eq. \eqref{eq:LR_eq} 
within a singles and doubles parameterization are 
\begin{align}
\label{eq:term1}
    & \left<0\left|\left[\hat{E}_{\nu_k\nu_c}\hat{E}_{\nu_l\nu_d}, \left[\sum_{pqrs}g_{pqrs}\hat{e}_{pqrs}, \hat{E}_{\nu_a\nu_i}\hat{E}_{\nu_b\nu_j}\right]\right]\right|0\right> \\
    \label{eq:term2}
    & \left<0\left|\left[\hat{E}_{\nu_k\nu_c}\hat{E}_{\nu_l\nu_d}, \left[\sum_{pqrs}g_{pqrs}\hat{e}_{pqrs}, \hat{E}_{\nu_i\nu_a}\hat{E}_{\nu_j\nu_b}\right]\right]\right|0\right>.
\end{align}
They are both matrix elements of the Hessian, obtained by inserting Eq. \eqref{eq:R} in their respective matrix elements. The first, Eq. \eqref{eq:term1}, is obtained by inserting Eq. \eqref{eq:R} in $\braket{ 0 | \left[ \hat G^\dagger_n \left[ \hat H, \hat G_m \right]\right] | 0 }$ of Eq. \eqref{eq:LR_A} and the second, Eq. \eqref{eq:term2}, is obtained by inserting Eq. \eqref{eq:R} in $\braket{ 0 | \left[ \hat G^\dagger_n \left[ \hat H, \hat G^\dagger_m \right]\right] | 0 }$ of Eq. \eqref{eq:LR_B}. Due to rank reduction \cite{Helgaker2013book}, these matrix elements will contain at most the 4-electron reduced density matrix (4-RDM). 

As the terms in 
Eqs.~\eqref{eq:term1} and \eqref{eq:term2}
are the largest elements in the naive qLRSD approach, we need to consider all RDMs up to four particles (4-RDM) to be able to formulate naive qLRSD in terms of reduced density matrices. These are given in second quantization as

\begin{align}
    \Gamma^{[1]}_{pq} = & \sum_{\tau\in\{\alpha,\beta\}} \left<0\left|\hat a^\dagger_{p\tau}\hat a_{q\tau}\right|0\right> = \left<0\left|\hat{E}_{pq}\right|0\right> \\
    \Gamma^{[2]}_{pqrs} = & \sum_{\tau\gamma\in\{\alpha,\beta\}} \left<0\left|\hat a^\dagger_{p\tau}\hat a^\dagger_{r\gamma}\hat a_{s\gamma}\hat a_{q\tau}\right|0\right> = \left<0\left|\hat{e}_{pqrs}\right|0\right> \\
    \Gamma^{[3]}_{pqrstu} = & \sum_{\tau\gamma\delta\in\{\alpha,\beta\}} \left<0\left|\hat a^\dagger_{p\tau}\hat a^\dagger_{r\gamma}\hat a^\dagger_{t\delta}\hat a_{u\delta}\hat a_{s\gamma}\hat a_{q\tau}\right|0\right> \\\nonumber 
    = & \left<0\left|\hat{E}_{pq}\hat{E}_{rs}\hat{E}_{tu}\right|0\right> - \delta_{ts}\Gamma^{[2]}_{pqru} - \delta_{rq}\Gamma^{[2]}_{pstu} - \delta_{tq}\Gamma^{[2]}_{purs} - \delta_{ts}\delta_{rq}\Gamma^{[1]}_{pu} \\
    \Gamma^{[4]}_{pqrstumn} = & \sum_{\tau\gamma\delta\sigma\in\{\alpha,\beta\}} \left<0\left|\hat a^\dagger_{p\tau}\hat a^\dagger_{r\gamma}\hat a^\dagger_{t\delta}\hat a^\dagger_{m\sigma}\hat a_{n\sigma}\hat a_{u\delta}\hat a_{s\gamma}\hat a_{q\tau}\right|0\right> \\\nonumber 
    = & \left<0\left|\hat{E}_{pq}\hat{E}_{rs}\hat{E}_{tu}\hat{E}_{mn}\right|0\right> - \delta_{rq}\Gamma^{[3]}_{pstumn} - \delta_{tq}\Gamma^{[3]}_{pursmn} - \delta_{mq}\Gamma^{[3]}_{pnrstu} - \delta_{mu}\Gamma^{[3]}_{pqrstn} 
    \\\nonumber
    & - \delta_{ts}\Gamma^{[3]}_{pqrumn} - \delta_{ms}\Gamma^{[3]}_{pqrntu} - \delta_{mu}\delta_{rq}\Gamma^{[2]}_{pstn} - \delta_{mu}\delta_{tq}\Gamma^{[2]}_{pnrs} - \delta_{ts}\delta_{mu}\Gamma^{[2]}_{pqrn} \\\nonumber
    & - \delta_{ts}\delta_{rq}\Gamma^{[2]}_{pumn} - \delta_{ts}\delta_{mq}\Gamma^{[2]}_{pnru} - \delta_{ms}\delta_{rq}\Gamma^{[2]}_{pntu} - \delta_{ms}\delta_{tq}\Gamma^{[2]}_{purn} - \delta_{mu}\delta_{ts}\delta_{rq}\Gamma^{[1]}_{pn}
\end{align}
where $a^\dagger$ and $a$ are the creation and annihilation operators.

Using the separation of the wave function into inactive, active, and virtual components in Eq. \eqref{eq:AS_seperation}, the RDMs may be expressed by lower order RDMs when not all indices are in the active space and equals zero when any virtual indices are present. Crucially, inactive indices trivially reduce to Kronecker deltas and thus we only need to calculate the RDMs with active space indices on quantum hardware. Moreover, since the RDMs fulfill the symmetries shown in Eqs. \eqref{SI-eq:rdm1_sym} to \eqref{SI-eq:rdm4_sym} in the SI 
(Section \ref{SI-app:sym}), 
it is only necessary to consider the non-redundant combinations of inactive and active indices as shown in Eqs. \eqref{eq:rdm1_expression} to \eqref{eq:rdm4_expression} reducing the number of elements that need to be measured on the quantum hardware by a factor equal to the number of symmetries. For the 4-RDM there are 48 symmetries, and so only $\frac{1}{48}$ of the elements need to be measured.
\begin{equation}
    \Gamma^{[1]}_{pq} = \left\{\begin{array}{ll}
        2\delta_{pq} & pq \in II \\
        \left<0\left|\hat{E}_{pq}\right|0\right> & pq \in AA \\
        0 & \text{otherwise} \\
\end{array} \right. \label{eq:rdm1_expression}
\end{equation}
\begin{equation}
    \Gamma^{[2]}_{pqrs} = \left\{\begin{array}{ll}
        4\delta_{pq}\delta_{rs} - 2\delta_{rq}\delta_{ps} & pqrs \in IIII \\
        2\delta_{rs} \Gamma^{[1]}_{pq} & pqrs \in AAII \\
        - \delta_{ps}\Gamma^{[1]}_{rq} & pqrs \in IAAI \\
        \left<0\left|\hat{e}_{pqrs}\right|0\right> & pqrs \in AAAA \\
        0 & \text{otherwise} \\
\end{array} \right. \label{eq:rdm2_expression}
\end{equation}
\begin{equation}
    \Gamma^{[3]}_{pqrstu} = \left\{\begin{array}{ll}
        2\delta_{pq}\Gamma^{[2]}_{rstu} - \delta_{rq}\Gamma^{[2]}_{pstu} - \delta_{tq}\Gamma^{[2]}_{purs} & pqrstu \in IIIIII \\
        4 \delta_{rs}\delta_{tu}\Gamma^{[1]}_{pq} - 2\delta_{ts}\delta_{ru}\Gamma^{[1]}_{pq} & pqrstu \in AAIIII \\
        \delta_{pu}\delta_{ts}\Gamma^{[1]}_{rq} - 2\delta_{ps}\delta_{tu}\Gamma^{[1]}_{rq} & pqrstu \in IAAIII \\
        2\delta_{tu}\Gamma^{[2]}_{pqrs} & pqrstu \in AAAAII \\
        - \delta_{ru}\Gamma^{[2]}_{pqts} & pqrstu \in AAIAAI \\
        \left<0\left|\hat{E}^{prt}_{qsu}\right|0\right> & pqrstu \in AAAAAA \\
        0 & \text{otherwise} \\
\end{array} \right. \label{eq:rdm3_expression}
\end{equation}
\begin{equation}
    \Gamma^{[4]}_{pqrstumn} = \left\{\begin{array}{ll}
        2\delta_{pq}\Gamma^{[3]}_{rstumn} - \delta_{rq}\Gamma^{[3]}_{pstumn} - \delta_{tq}\Gamma^{[3]}_{pursmn} - \delta_{mq}\Gamma^{[3]}_{pnrstu} & pqrstumn \in IIIIIIII \\
        \Gamma^{[1]}_{pq}\Gamma^{[3]}_{rstumn} & pqrstumn \in AAIIIIII \\
        - \frac{1}{2}\Gamma^{[1]}_{rq}\Gamma^{[3]}_{pstumn} & pqrstumn \in IAAIIIII \\
        \left( 4\delta_{tu}\delta_{mn} - 2\delta_{mu}\delta_{tn} \right)\Gamma^{[2]}_{pqrs} & pqrstumn \in AAAAIIII \\
        \left( \delta_{rn}\delta_{mu} - 2\delta_{ru}\delta_{mn} \right)\Gamma^{[2]}_{pqts} & pqrstumn \in AAIAAIII \\
        \delta_{ps}\delta_{tn}\Gamma^{[2]}_{rqmu} + \delta_{pn}\delta_{ts}\Gamma^{[2]}_{rumq} & pqrstumn \in IAAIIAAI \\
        2\delta_{mn} \Gamma^{[3]}_{pqrstu} & pqrstumn \in AAAAAAII \\
        - \delta_{tn}\Gamma^{[3]}_{pqrsmu} & pqrstumn \in AAAAIAAI \\
        \left<0\left|\hat{E}^{prtm}_{qsun}\right|0\right> & pqrstumn \in AAAAAAAA \\
        0 & \text{otherwise} \\
\end{array} \right.. \label{eq:rdm4_expression}
\end{equation}
We have also only included one permutation of the unique combinations of inactive-active pairs. As an example, let us take the term $\Gamma^{[2]}_{pqrs} = - \delta_{ps}\Gamma^{[1]}_{rq}$ with $pqrs \in IAAI$. For this combination of inactive and active orbitals, there are two possible permutations
\begin{align}
    \Gamma^{[2]}_{pqrs} = - \delta_{ps}\Gamma^{[1]}_{rq} \quad & pqrs \in IAAI \\
    \Gamma^{[2]}_{pqrs} = - \delta_{rq}\Gamma^{[1]}_{ps} \quad & pqrs \in AIIA~,
\end{align}
where the permutations on the right hand side of the equations have been moved according to the permutation in the inactive and active indices $(pq \leftrightarrow rs)$. Working equations for all matrix elements needed to solve the linear response equation can be found in Section \ref{SI-app:working_equations} in the SI.

\subsection{Scaling}
\label{ssec:scaling}

In terms of the number of elements, the three terms with the largest scaling are
\begin{align}
    \left<0\left|\left[\hat G,\left[\hat H,\hat G\right]\right]\right|0\right> & \rightarrow N_{A_\text{occ}}^4N_{A_\text{unocc}}^4N_{I+A}^2 \label{eq:scaling_GG}\\
    \left<0\left|\left[\hat q,\left[\hat H,\hat G\right]\right]\right|0\right> & \rightarrow N_{A_\text{occ}}^2N_{A_\text{unocc}}^2 \max
\{N_{I+A}^2N_IN_A, N_\text{orb}N_{I+A}N_IN_V, N_\text{orb}N_{I+A}N_AN_V\} \\
    \left<0\left|\left[\hat q,\left[\hat H,\hat q\right]\right]\right|0\right> & \rightarrow 
    \max
\{N_{I+A}^2N_I^2N_A^2, N_\text{orb}^2N_I^2N_V^2, N_\text{orb}^2N_A^2N_V^2\}. \label{eq:scaling_qq}
\end{align}
In the equations above, various indices have been used to refer to the number of orbitals in different spaces. These are the number of orbitals in the inactive space $(N_I)$, active space $(N_A)$, virtual space $(N_V)$, active space that is inactive in the Hartree-Fock reference state $(N_{A_\text{occ}})$, active space that is virtual in the Hartree-Fock reference state $(N_{A_\text{unocc}})$, inactive and active space $(N_{I+A})$, and in total $(N_\text{orb})$.
For a minimal basis set Eq.~\eqref{eq:scaling_GG} will be the dominating term. However, with an increasing basis set size the virtual indices will dominate the scaling eventually causing Eq. \eqref{eq:scaling_qq} to become the dominant term.
Regarding memory usage, the two-electron integrals and the 4-RDM are of interest as they scale as $N_\text{orb}^4$ and $N_A^8$, respectively. Since we construct the entire Hessian,  it is also important to consider its size
for larger systems, as it scales
\begin{equation}
    \label{eq:hessian_size}
    \textbf{E}^{[2]} \rightarrow (2(N_{\hat{q}} + N_{\hat{G}}))^2 = 4(N_{\hat{q}} + N_{\hat{G}})^2
\end{equation}
where $N_{\hat q}$ and $N_{\hat G}$ are the number of orbital rotation and active space excitation operators. $N_{\hat G}$ scale with the active space size whereas $N_{\hat q}$ scale with the size of the inactive, active, and virtual spaces. To reduce the impact of the scaling of the Hessian it is possible to use a subspace approach instead of constructing the full Hessian explicitly \cite{reinholdt2024subspace}. This is a topic for future research.

\section{\label{sec:comp}Computational Details}

As the oo-UCC wave function is decomposed into an inactive, active, and virtual space we here adopt the CAS-like notation 
oo-UCC($n$, $\sigma$), 
where $n$ is the number of electrons in the active space and $\sigma$ is the number of orbitals in the active space. The excitation ranks included in the wave function's active space is truncated at the singles and doubles level. Meaning, oo-UCCSD(6,6) uses a (6,6) active space with cluster operators to the level of singles and doubles excitations in the active space.

The oo-UCC calculations were performed using our in-house quantum chemistry software \verb|SlowQuant| \cite{slowquant} which is interfaced with \verb|PySCF| \cite{Sun2015,Sun2020} for one and two-electron integrals as well as the initial MP2 optimization for the MP2 natural orbitals, and with \verb|Qiskit-aer| \cite{Qiskit} as the quantum simulation backend. 
All calculations used the MP2 natural orbitals as their initial guess. After the oo-UCC optimization, the RDMs in the active space were calculated using \verb|SlowQuant| and then 
fed into
our standalone \verb|DensityMatrixDrivenModule| (\verb|DMDM|) for the calculation of the complete space RDMs according to Eqs. \eqref{eq:rdm1_expression} to \eqref{eq:rdm4_expression} and, subsequently, of the 
linear response properties, specifically excitation energies and spectral strengths. The classic CASSCF reference calculations were performed using \verb|Dalton| \cite{dalton} and MP2 natural orbitals as the initial guess.

To show the performance capabilities of using RDMs for naive qLRSD, excitation energies and oscillator strengths were calculated for benzene 
and the excitation energies, oscillator strengths, and rotational strengths of $R$-methyloxirane were calculated using active space RDMs from an oo-UCCSD(6,6)/ cc-pVTZ \cite{dunning1989a} calculation in \verb|SlowQuant|. The absorption spectrum of benzene and the ECD spectrum of $R$-methyloxirane were then compared to CASSCF(6,6)/cc-pVTZ results obtained from \verb|Dalton|. 

Finally, a shot noise analysis was performed. For this, 10 samples of the absorption spectrum of water were calculated on top of a noise-free oo-UCCSD(4,4)/cc-pVTZ wave function using 10k, 50k, and 100k shots. The active space RDMs were simulated through \verb|SlowQuant| using \verb|Qiskit-aer| as the quantum simulator backend. The wave function on the quantum simulator used a UCCSD ansatz with Parity mapping.
The reference noiseless absorption spectrum 
was obtained using \verb|SlowQuant|.

All geometries used are available in Section \ref{SI-sec:geom} of the SI, and examples of code calculating excitation energies, oscillator strengths, and rotational strengths using \verb|DMDM| are available as well in Section \ref{SI-sec:code_examples} of the SI. All spectra use a Lorentzian convolution with FWHM broadening of 0.2 eV and compare the first 399 excitation energies as this is the maximum that \verb|Dalton| allows by default.
Note that this exceedingly large number of states is used with the sole purpose of comparing CASSCF with oo-UCCSD on a broad spectral range; not all states are expected to be physically significant.

\section{\label{sec:results}Results and Discussion}

Having derived working equations for an RDM formulation of naive qLRSD, we will in the following section compare its results against CASSCF(6,6) LR for the calculation of excitation energies, oscillator strengths, and rotational strengths. For this, we provide 
calculated absorption spectra of benzene in Sec.~\ref{ssec:benzene} and ECD spectra of $R$-methyloxirane in Sec. \ref{ssec:methyloxirane}. 
Finally, we present an analysis of our formulation's 
resilience
to shot noise in 
Sec.~\ref{ssec:shot_noise}.

\subsection{Benzene}
\label{ssec:benzene}

\begin{figure}[H]
    \centering
    \includegraphics[width=.7\textwidth]{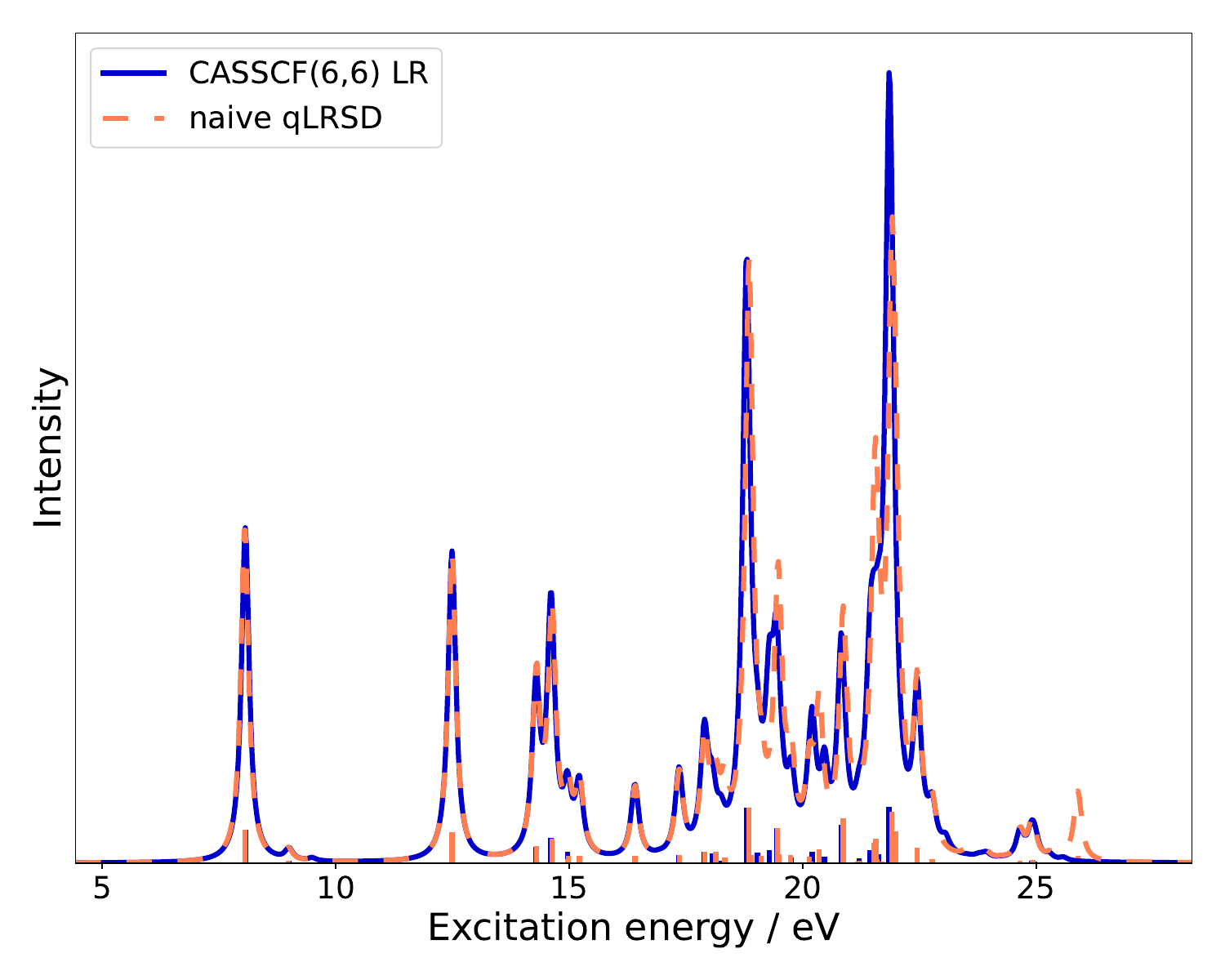}
    \caption{Benzene. Absorption spectrum of the lowest 399 excitation energies using CASSCF(6,6) LR (blue), and qLRSD (dashed orange) based on the oo-UCCSD(6,6) wave function using the cc-pVTZ basis set.}
    \label{fig:absorption}
\end{figure}

In Fig.~\ref{fig:absorption}, we show the absorption spectra of benzene using CASSCF(6,6) LR (blue), and qLRSD (dashed orange). Note that the CASSCF(6,6) LR calculation includes the complete active excitation space whereas qLRSD only includes singles and doubles excitations in the active excitation space, leading to a reduction in active space parameters from 174 for CASSCF(6,6) LR to 54 for qLRSD. The reduced excitation space does not result in a significant change in the absorption spectrum, with most peaks being identical in both excitation energy and oscillator strength. The minor deviations between the two spectra are caused by the truncated active excitation space in the linear response. To gain a more accurate description for these excitations, a higher active excitation space is needed \cite{ziems2023options}.

\subsection{\textit{R}-methyloxirane}
\label{ssec:methyloxirane}

\begin{figure}[H]
    \centering
    \includegraphics[width=.7\textwidth]{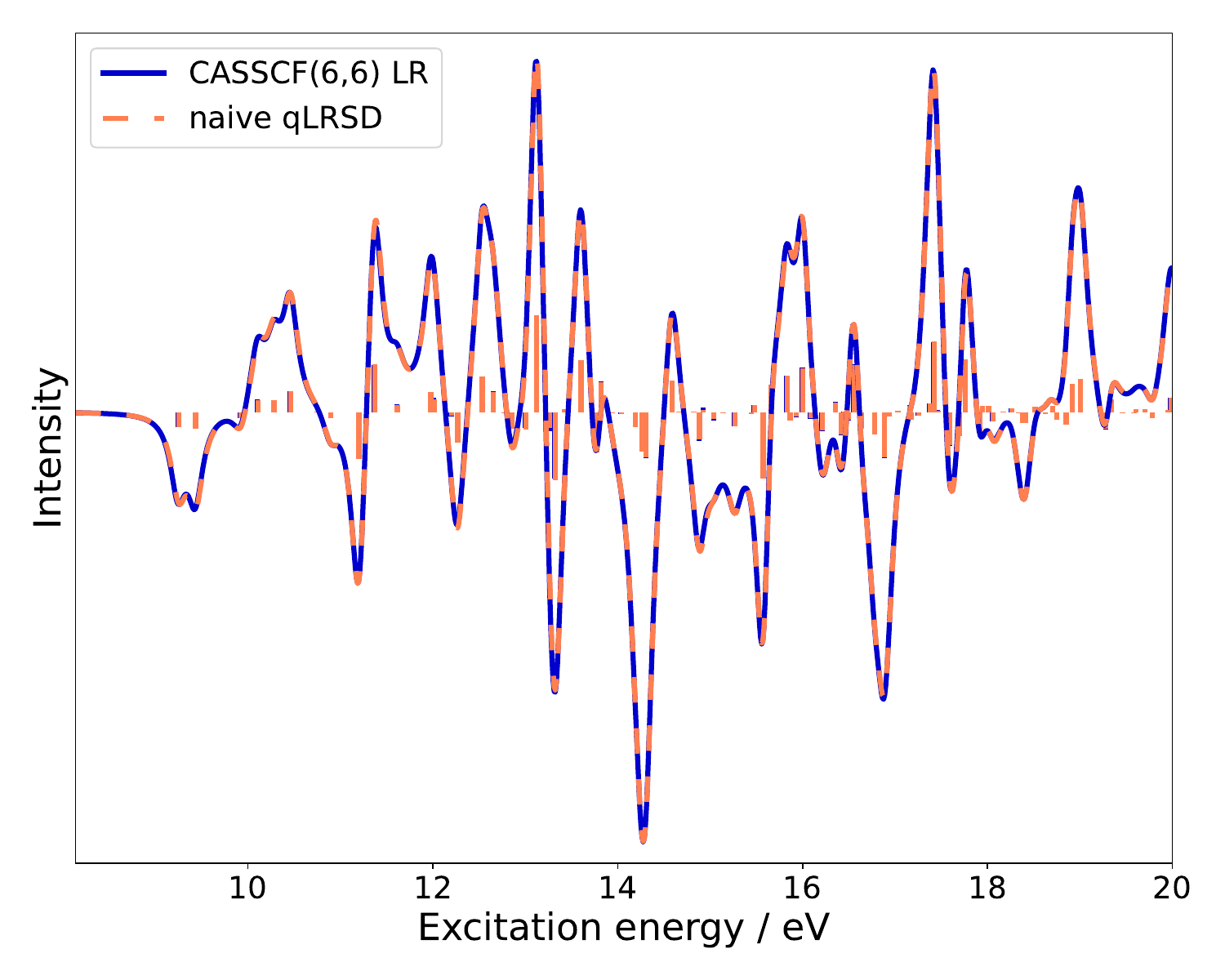}
    \caption{$R$-methyloxirane. ECD spectrum of the lowest 97 excitation energies using CASSCF(6,6) LR (blue), and qLRSD (dashed orange) based on the oo-UCCSD(6,6) wave function using the cc-pVTZ basis set.}
    \label{fig:ecd}
\end{figure}

The ECD spectra of $R$-methyloxirane using CASSCF(6,6) LR and qLRSD are shown in Fig. \ref{fig:ecd}. 
In the SI, the ECD spectrum containing all 399 calculated excitation energies (Fig. \ref{SI-fig:ecd_full}) and the absorption spectrum (\ref{SI-fig:absorption_methyloxirane}) can be found (Section \ref{SI-sec:TabFig}).

The truncated active excitation space of qLRSD leads to the same parameter reduction as for benzene, whilst no discernible difference between the CASSCF(6,6) and the qLRSD spectra is visible. 
Therefore, the truncated excitation space is sufficient in describing the excitation energies and rotational strengths of $R$-methyloxirane. 
As can be appreciated from Fig. \ref{SI-fig:absorption_methyloxirane}, this also applies (not surprisingly) to the oscillator strengths of $R$-methyloxirane where truncating the excitation space to singles and doubles 
proves sufficient for 
oscillator strengths of comparable quality to CASSCF(6,6).

\subsection{Shot noise analysis}
\label{ssec:shot_noise}

\begin{figure}[H]
    \centering
    \includegraphics[width=.49\textwidth]{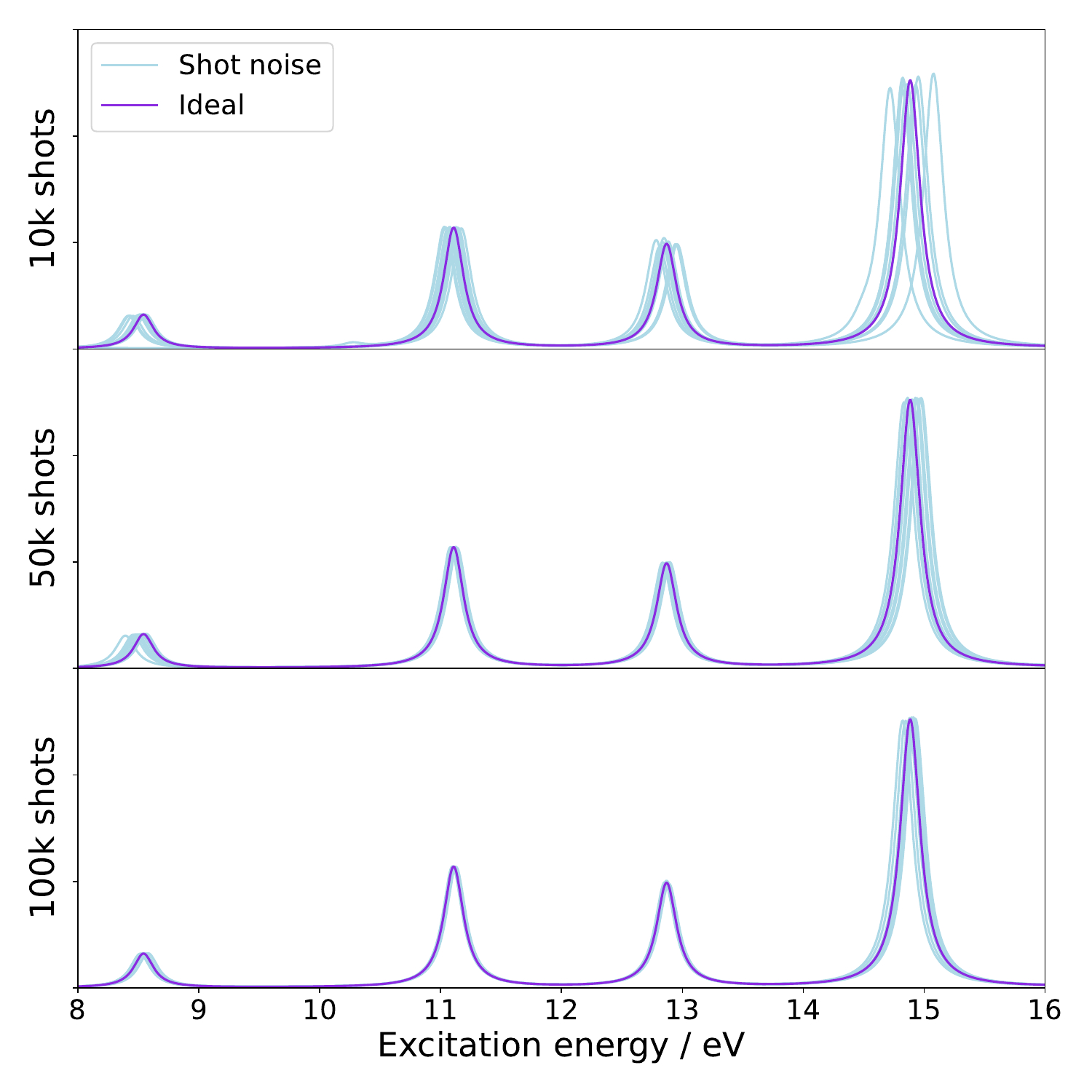}
    \includegraphics[width=.49\textwidth]{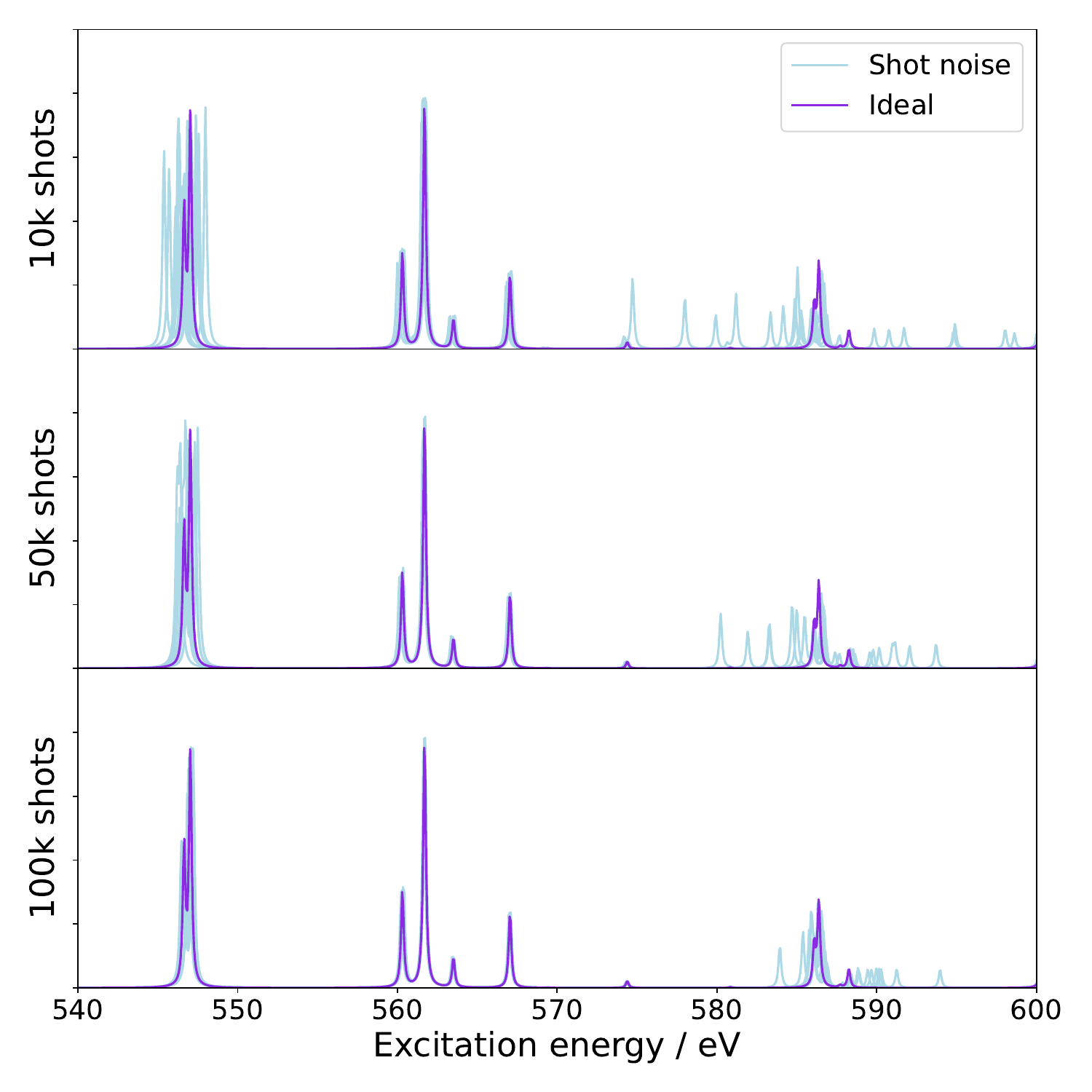}
    \caption{Water. Absorption spectrum in the low energy region of 8 to 16 eV (valence excitations, left) and in the high energy region of 540 to 600 eV (oxygen K-edge, right) on simulated hardware to model shot noise. Top panels: 10\,000 shots; middle panels: 50\,000 shots; bottom panels: 100\,000 shots. Each panel contains ten spectra including shot noise and one `ideal' spectrum based on a state-vector calculation.}
    \label{fig:noise_low_energy}
\end{figure}

In Fig. \ref{fig:noise_low_energy}, the absorption spectra of H$_2$O(4,4)/cc-pVTZ with shot noise from a quantum simulator can be found. We show, respectively, part of the valence excitation energy region (8 to 16 eV) and around the oxygen K-edge (starting at 540 eV) using qLRSD. Exact values for variance and standard deviations in the two energy regions can be found in the SI in table \ref{SI-tab:shot_noise} (Section \ref{SI-sec:TabFig}). It is evident that a higher amount of shots leads to more reliable results. Moreover, the peaks show differences in their convergence to the ideal spectrum with increasing number of shots. Looking at individual peaks, the first three peaks in the low energy region and the peaks between 560 and 570 eV of Fig. \ref{fig:noise_low_energy} have low errors when comparing the simulated spectra with shot noise to the ideal spectrum. The remaining peaks converge toward the ideal result but still show some deviations like the fourth peak in the low energy region and the first peak in the high energy region of Fig. \ref{fig:noise_low_energy}. Significant variance in the region $>$580 eV of Fig. \ref{fig:noise_low_energy} also appears. A more in-depth analysis of the excitation energies and oscillator strengths to identify why some peaks have larger variance and standard deviation is a topic for future research. 

\section{\label{sec:summary}Summary}

In this paper, we have formulated and implemented a density-matrix-driven approach to our recent naive qLRSD approach, that drastically reduces the classical computational costs. This will allow for more demanding calculations employing larger basis sets and on larger molecules. To illustrate our approach, we reported the absorption spectrum of benzene and the ECD spectrum of 
$R$-methyloxirane calculated in a (6,6) active space using the cc-pVTZ basis set. 
These constitute the largest basis sets employed so far for quantum approaches to molecular properties on moderate sized molecules and is an important step to achieving meaningful results on NISQ-era quantum devices.
The truncation of the excitation rank in the active space leads to a significant reduction in the number of active space parameters with results almost indistinguishable from classic CASSCF simulations with a complete active space. 

Using the active space approximation allows us to separate the RDMs in the complete space into a sum of RDMs over different indices in the inactive and active spaces, where only the latter necessitates evaluation on quantum device. Crucially, when not all indices are in the active space, the order of the RDMs can be significantly reduced. Moreover, the symmetries of the RDMs leads to a reduction of the number of elements in the RDMs that need to be measured. Depending on the number of symmetries in the different order of RDMs, this can lead to a reduction by a factor of up to 48 for the 4-RDM.

Additionally, we provide proof-of-principle quantum simulations with shot noise from a simulated quantum backend. In the analysis of shot noise, we observe the trivial trend that more shots converge to the ideal spectrum. However, some peaks converge quickly, while others have larger variances. The reason behind this observation is a point for further investigation. We are currently investigating the impact of using a cumulant-based density matrix approach where higher-order RDMs, such as the 3- and 4-electron RDMs, are approximated from lower order RDMs \cite{Harris2002,Zgid2009,Saitow2013}. This would lead to a further reduction in classic computational costs and in quantum measurements.

\section*{Acknowledgments}
We acknowledge support from the Novo Nordisk Foundation (NNF) for the focused research project \textit{Hybrid Quantum Chemistry on Hybrid Quantum Computers} (HQC)$^2$ (grant number: NNFSA220080996).



\newpage
\bibliography{literature}

\end{document}


\date{April 25, 2024}

\maketitle

\clearpage

\section{Symmetries}
\label{app:sym}

The symmetries of the RDMs are given below
\begin{align}
    \Gamma^{[1]}_{pq} = & \Gamma^{[1]}_{qp} \label{eq:rdm1_sym} \\
    \Gamma^{[2]}_{pqrs} = & \Gamma^{[2]}_{rspq} = \Gamma^{[2]}_{qpsr} = \Gamma^{[2]}_{srqp} \label{eq:rdm2_sym} \\
    \Gamma^{[3]}_{pqrstu} = & \Gamma^{[3]}_{pqturs} = \Gamma^{[3]}_{rspqtu} = \Gamma^{[3]}_{rstupq} = \Gamma^{[3]}_{tupqrs} = \Gamma^{[3]}_{turspq} \label{eq:rdm3_sym} \\\nonumber
    = & \Gamma^{[3]}_{qpsrut} = \Gamma^{[3]}_{qputsr} = \Gamma^{[3]}_{srqput} = \Gamma^{[3]}_{srutqp} = \Gamma^{[3]}_{utqpsr} = \Gamma^{[3]}_{utsrqp} \\
    \Gamma^{[4]}_{pqrstumn} = & \Gamma^{[4]}_{pqrsmntu} = \Gamma^{[4]}_{pqtursmn} = \Gamma^{[4]}_{pqtumnrs} = \Gamma^{[4]}_{pqmnrstu} = \Gamma^{[4]}_{pqmnturs}\label{eq:rdm4_sym} \\\nonumber
    = & \Gamma^{[4]}_{rspqtumn} = \Gamma^{[4]}_{rspqmntu} = \Gamma^{[4]}_{rstupqmn} = \Gamma^{[4]}_{rstumnpq} = \Gamma^{[4]}_{rsmnpqtu} = \Gamma^{[4]}_{rsmntupq} \\\nonumber
    = & \Gamma^{[4]}_{tupqrsmn} = \Gamma^{[4]}_{tupqmnrs} = \Gamma^{[4]}_{turspqmn} = \Gamma^{[4]}_{tursmnpq} = \Gamma^{[4]}_{tumnpqrs} = \Gamma^{[4]}_{tumnrspq} \\\nonumber
    = & \Gamma^{[4]}_{mnpqrstu} = \Gamma^{[4]}_{mnpqturs} = \Gamma^{[4]}_{mnrspqtu} = \Gamma^{[4]}_{mnrstupq} = \Gamma^{[4]}_{mntupqrs} = \Gamma^{[4]}_{mnturspq} \\\nonumber
    = & \Gamma^{[4]}_{qpsrutnm} = \Gamma^{[4]}_{qpsrnmut} = \Gamma^{[4]}_{qputsrnm} = \Gamma^{[4]}_{qputnmsr} = \Gamma^{[4]}_{qpnmsrut} = \Gamma^{[4]}_{qpnmutsr} \\\nonumber
    = & \Gamma^{[4]}_{srqputnm} = \Gamma^{[4]}_{srqpnmut} = \Gamma^{[4]}_{srutqpnm} = \Gamma^{[4]}_{srutnmqp} = \Gamma^{[4]}_{srnmqput} = \Gamma^{[4]}_{srnmutqp} \\\nonumber
    = & \Gamma^{[4]}_{utqpsrnm} = \Gamma^{[4]}_{utqpnmsr} = \Gamma^{[4]}_{utsrqpnm} = \Gamma^{[4]}_{utsrnmqp} = \Gamma^{[4]}_{utnmqpsr} = \Gamma^{[4]}_{utnmsrqp} \\\nonumber
    = & \Gamma^{[4]}_{nmqpsrut} = \Gamma^{[4]}_{nmqputsr} = \Gamma^{[4]}_{nmsrqput} = \Gamma^{[4]}_{nmsrutqp} = \Gamma^{[4]}_{nmutqpsr} = \Gamma^{[4]}_{nmutsrqp}.
\end{align}

\section{Working equations}
\label{app:working_equations}

The key equations of linear reponse are the linear response equation, Eq. \eqref{main-eq:LR_eq}, and its homogeneous eigenvalue equation, Eq. \eqref{main-eq:LR_exc}.
To solve them, the Hessian and the Hermitian metric must be calculated first. The Hessian and the Hermitian metric contain the sub-matrices in Eqs. \eqref{main-eq:LR_A} to \eqref{main-eq:LR_delta}. With the definition of the orbital rotation operators $\hat q_\mu$ and $\hat G_n$ in Eqs. \eqref{main-eq:Q} and \eqref{main-eq:R}, these may be expressed by the matrix elements in Eqs. \eqref{eq:E_hE} to \eqref{eq:EE_EE} as RDMs. All indices used in this section should be read as general indices.

\begin{align}\label{eq:E_hE}
    &\left<0\left|\left[\hat{E}_{mn}, \sum_{pq}h_{pq}\hat{E}_{pq}\right]\right|0\right>\\\nonumber
    &= \sum_p h_{np}\Gamma^{[1]}_{mp} - \sum_p h_{pm}\Gamma^{[1]}_{pn}
\end{align}

\begin{align}\label{eq:E_ge}
     &\left<0\left|\left[\hat{E}_{mn}, \sum_{pqrs}g_{pqrs}\hat{e}_{pqrs}\right]\right|0\right>\\\nonumber
    &= 2\sum_{pqr}g_{npqr}\Gamma^{[2]}_{mpqr} - 2\sum_{pqr}g_{pmqr}\Gamma^{[2]}_{pnqr}
\end{align}

\begin{align}\label{eq:EE_hE}
    &\left<0\left|\left[\hat{E}_{tu}\hat{E}_{mn}, \sum_{pq}h_{pq}\hat{E}_{pq}\right]\right|0\right>\\\nonumber
    & = \sum_q h_{uq}\Gamma^{(2)}_{tqmn} - \sum_p h_{pt} \big( \Gamma^{(2)}_{pumn} - \delta_{mu}\Gamma^{(1)}_{pn} \big) \\\nonumber
    &\quad + \sum_q h_{nq} \big( \Gamma^{(2)}_{tumq} + \delta_{mu}\Gamma^{(1)}_{tq} \big) - \sum_p h_{pm}\Gamma^{(2)}_{tupn}
\end{align}

\begin{align}\label{eq:EE_ge}
    & \left<0\left|\left[\hat{E}_{tu}\hat{E}_{mn}, \sum_{pqrs}g_{pqrs}\hat{e}_{pqrs}\right]\right|0\right> \\\nonumber
    & = 2 \bigg( \sum_{pqr}g_{npqr}\Gamma^{[3]}_{tumpqr} + \sum_{pqr}g_{upqr}\Gamma^{[3]}_{mntpqr} - \sum_{pqr}g_{pmqr}\Gamma^{[3]}_{tupnqr} - \sum_{pqr}g_{ptqr}\Gamma^{[3]}_{mnpuqr} \\
    &\quad + \sum_{pqr}\delta_{mu}g_{pqnr}\Gamma^{[2]}_{trpq} - \sum_{pqr}\delta_{mu}g_{ptqr}\Gamma^{[2]}_{pnqr} + \sum_{pq}g_{upnq}\Gamma^{[2]}_{tpmq} - \sum_{pq}g_{pmqt}\Gamma^{[2]}_{pnqu} \bigg)
\end{align}

\begin{align}\label{eq:E_hE_E}
     &\left<0\left|\left[\hat{E}_{tu}, \left[\sum_{pq}h_{pq}\hat{E}_{pq}, \hat{E}_{mn}\right]\right]\right|0\right>\\\nonumber
    &= -\sum_p h_{np}\delta_{mu}\Gamma^{[1]}_{tp} + h_{nt}\Gamma^{[1]}_{mu} + h_{um}\Gamma^{[1]}_{tn} - \sum_p h_{pm}\delta_{tn}\Gamma^{[1]}_{pu}
\end{align}

\begin{align}\label{eq:E_ge_E}
    &\left<0\left|\left[\hat{E}_{tu}, \left[\sum_{pqrs}g_{pqrs}\hat{e}_{pqrs}, \hat{E}_{mn}\right]\right]\right|0\right>\\\nonumber
    &= - 2\sum_{pqr}g_{npqr}\delta_{mu}\Gamma^{[2]}_{tpqr} + 2\sum_{pq}g_{ntpq}\Gamma^{[2]}_{mupq} - 2\sum_{pq}g_{npuq}\Gamma^{[2]}_{mptq} + 2\sum_{pq}g_{npqt}\Gamma^{[2]}_{mpqu} \\\nonumber
    &\quad + 2\sum_{pq}g_{umpq}\Gamma^{[2]}_{tnpq} - 2\sum_{pqr}g_{pmqr}\delta_{tn}\Gamma^{[2]}_{puqr} + 2\sum_{pq}g_{pmuq}\Gamma^{[2]}_{pntq} - 2\sum_{pq}g_{pmqt}\Gamma^{[2]}_{pnqu}
\end{align}

\begin{align}\label{eq:EE_hE_E}
    & \left<0\left|\left[\hat{E}_{ij}\hat{E}_{kl}, \left[\sum_{pq}h_{pq}\hat{E}_{pq}, \hat{E}_{tu}\right]\right]\right|0\right> \\\nonumber
    & = h_{ui}\Gamma^{[2]}_{tjkl} + h_{uk}\Gamma^{[2]}_{ijtl} + h_{jk}\Gamma^{[2]}_{iukl} + h_{lt}\Gamma^{[2]}_{ijku} + \delta_{kj}h_{ui}\Gamma^{[1]}_{tl} + \delta_{kj}h_{lt}\Gamma^{[1]}_{iu} \\\nonumber
    &\quad - \sum_p\delta_{tj}h_{up}\Gamma^{[2]}_{ipkl} - \sum_p\delta_{tl}h_{up}\Gamma^{[2]}_{ijkp} - \sum_p\delta_{iu}h_{pt}\Gamma^{[2]}_{pjkl} - \sum_p\delta_{ku}h_{pt}\Gamma^{[2]}_{ijpl} \\\nonumber
    &\quad - \sum_p\delta_{tl}\delta_{kj}h_{up}\Gamma^{[1]}_{ip} - \sum_p\delta_{iu}\delta_{kj}h_{up}\Gamma^{[1]}_{pl}
\end{align}

\begin{align}\label{eq:EE_ge_E}
    & \left<0\left|\left[\hat{E}_{ij}\hat{E}_{kl}, \left[\sum_{pqrs}g_{pqrs}\hat{e}_{pqrs}, \hat{E}_{tu}\right]\right]\right|0\right> \\\nonumber
    & = 2\bigg( \sum_pg_{jtlp}\Gamma^{[2]}_{iukp} + \sum_pg_{ltjp}\Gamma^{[2]}_{ipku} + \sum_pg_{uipk}\Gamma^{[2]}_{tjpl} + \sum_pg_{piuk}\Gamma^{[2]}_{pjtl} + \sum_{pq}g_{ukpq}\Gamma^{[3]}_{ijtlpq} \\\nonumber
    &\quad + \sum_{pq}g_{upqk}\Gamma^{[3]}_{ijtpql} + \sum_{pq}g_{uipq}\Gamma^{[3]}_{tjpqkl} + \sum_{pq}g_{upqi}\Gamma^{[3]}_{tpqjkl} + \sum_{pq}g_{ptlq}\Gamma^{[3]}_{ijpukq} + \sum_{pq}g_{ltpq}\Gamma^{[3]}_{ijkupq} \\\nonumber
    &\quad + \sum_{pq}g_{jtpq}\Gamma^{[3]}_{iupqkl} + \sum_{pq}g_{ptjq}\Gamma^{[3]}_{puiqkl} -\sum_{pq}g_{upjq}\Gamma^{[3]}_{tpiqkl} - \sum_{pq}g_{ptqk}\Gamma^{[3]}_{ijpuql} - \sum_{pq}g_{ptqi}\Gamma^{[3]}_{puqjkl} \\\nonumber
    &\quad - \sum_{pq}g_{uplq}\Gamma^{[3]}_{ijtpkq} -
    \sum_{pq}\delta_{iu}g_{ptqk}\Gamma^{[2]}_{pjql} + \sum_{pq}\delta_{kj}g_{uipq}\Gamma^{[2]}_{tlpq} + \sum_{pq}\delta_{kj}g_{upqi}\Gamma^{[2]}_{tpql} + \sum_{pq}\delta_{kj}g_{ptlq}\Gamma^{[2]}_{iqpu} \\\nonumber
    &\quad + \sum_{pq}\delta_{kj}g_{ltpq}\Gamma^{[2]}_{iupq} - \sum_{pq}\delta_{kj}g_{uplq}\Gamma^{[2]}_{iqtp} - \sum_{pq}\delta_{kj}g_{ptqi}\Gamma^{[2]}_{puql} - \sum_{pq}\delta_{ku}g_{ptqi}\Gamma^{[2]}_{plqj} - \sum_{pq}\delta_{tj}g_{uplq}\Gamma^{[2]}_{ipkq} \\\nonumber
    &\quad - \sum_{pq}\delta_{tl}g_{upjq}\Gamma^{[2]}_{iqkp} - \sum_{pqr}\delta_{iu}g_{ptqr}\Gamma^{[3]}_{pjqrkl} - \sum_{pqr}\delta_{ku}g_{ptqr}\Gamma^{[3]}_{ijplqr} - \sum_{pqr}\delta_{tj}g_{upqr}\Gamma^{[3]}_{ipqrkl} \\\nonumber
    &\quad - \sum_{pqr}\delta_{tl}g_{upqr}\Gamma^{[3]}_{ijlkpqr} - \sum_{pqr}\delta_{iu}\delta_{kj}g_{ptqr}\Gamma^{[2]}_{plqr} - \sum_{pqr}\delta_{tl}\delta_{kj}g_{upqr}\Gamma^{[2]}_{ipqr} \bigg)
\end{align}

\begin{align}\label{eq:E_hE_EE}
    & \left<0\left|\left[\hat{E}_{ij}, \left[\sum_{pq}h_{pq}\hat{E}_{pq}, \hat{E}_{tu}\hat{E}_{mn}\right]\right]\right|0\right> \\\nonumber
    & = h_{jt}\Gamma^{[2]}_{iumn} + h_{ui}\Gamma^{[2]}_{tjmn} + h_{jm}\Gamma^{[2]}_{tuin} + h_{ni}\Gamma^{[2]}_{tumj} + \delta_{mu}h_{jt}\Gamma^{[1]}_{in} + \delta_{mu}h_{ni}\Gamma^{[1]}_{tj} \\\nonumber
    &\quad + \sum_p\delta_{iu}h_{np}\Gamma^{[2]}_{tjmp} - \sum_p\delta_{iu}h_{pq}\Gamma^{[2]}_{pjmn} - \sum_p\delta_{iu}h_{pm}\Gamma^{[2]}_{tjpn} + \sum_p\delta_{mj}h_{pt}\Gamma^{[2]}_{puin} \\\nonumber
    &\quad - \sum_p\delta_{mj}h_{up}\Gamma^{[2]}_{tpin} - \sum_p\delta_{mj}h_{np}\Gamma^{[2]}_{tuip} + \sum_p\delta_{in}h_{up}\Gamma^{[2]}_{tpmj} - \sum_p\delta_{in}h_{pt}\Gamma^{[2]}_{pumj} \\\nonumber
    &\quad - \sum_p\delta_{in}h_{pm}\Gamma^{[2]}_{tupj} + \sum_p\delta_{tj}h_{pm}\Gamma^{[2]}_{iupn} - \sum_p\delta_{tj}h_{up}\Gamma^{[2]}_{ipmn} - \sum_p\delta_{tj}h_{np}\Gamma^{[2]}_{iump} \\\nonumber
    &\quad - \sum_p\delta_{in}\delta_{mu}h_{pt}\Gamma^{[1]}_{pj} - \sum_p\delta_{tj}\delta_{mu}h_{np}\Gamma^{[1]}_{ip}
\end{align}

\begin{align}\label{eq:E_ge_EE}
    & \left<0\left|\left[\hat{E}_{ij}, \left[\sum_{pqrs}g_{pqrs}\hat{e}_{pqrs}, \hat{E}_{tu}\hat{E}_{mn}\right]\right]\right|0\right> \\\nonumber
    & = 2\bigg( \sum_pg_{niup}\Gamma^{[2]}_{tpmj} + \sum_pg_{npui}\Gamma^{[2]}_{tjmp} + \sum_pg_{jtpm}\Gamma^{[2]}_{iupn} + \sum_pg_{ptjm}\Gamma^{[2]}_{puin} + \sum_{pq}g_{jmpq}\Gamma^{[3]}_{tuinpq} \\\nonumber
    &\quad + \sum_{pq}g_{pmjq}\Gamma^{[3]}_{tupniq} + \sum_{pq}g_{nipq}\Gamma^{[3]}_{tumjpq} + \sum_{pq}g_{npqi}\Gamma^{[3]}_{tumpqj} + \sum_{pq}g_{jtpq}\Gamma^{[3]}_{iupqmn} + \sum_{pq}g_{ptjq}\Gamma^{[3]}_{puiqmn} \\\nonumber
    &\quad + \sum_{pq}g_{uipq}\Gamma^{[3]}_{tjpqmn} + \sum_{pq}g_{upqi}\Gamma^{[3]}_{tpqjmn} - \sum_{pq}g_{piqm}\Gamma^{[3]}_{tupjqn} - \sum_{pq}g_{npjq}\Gamma^{[3]}_{tumpiq} - \sum_{pq}g_{ptqi}\Gamma^{[3]}_{puqjmn} \\\nonumber
    & \quad - \sum_{pq}g_{upjq}\Gamma^{[3]}_{tpiqmn} + \sum_{pq}\delta_{mu}g_{nipq}\Gamma^{[2]}_{tjpq} + \sum_{pq}\delta_{mu}g_{npqi}\Gamma^{[2]}_{tpqj} + \sum_{pq}\delta_{mu}g_{jtpq}\Gamma^{[2]}_{inpq} \\\nonumber
    &\quad + \sum_{pq}\delta_{mu}g_{ptjq}\Gamma^{[2]}_{pniq} - \sum_{pq}\delta_{iu}g_{ptqm}\Gamma^{[2]}_{pjqn} - \sum_{pq}\delta_{in}g_{ptqm}\Gamma^{[2]}_{puqj} - \sum_{pq}\delta_{mu}g_{npjq}\Gamma^{[2]}_{tpiq} \\\nonumber
    &\quad - \sum_{pq}\delta_{mu}g_{ptqi}\Gamma^{[2]}_{pnqj} - \sum_{pq}\delta_{mj}g_{npuq}\Gamma^{[2]}_{tqip} - \sum_{pq}\delta_{tj}g_{npuq}\Gamma^{[2]}_{iqmp} + \sum_{pqr}\delta_{iu}g_{npqr}\Gamma^{[3]}_{tjmpqr} \\\nonumber
    &\quad + \sum_{pqr}\delta_{in}g_{upqr}\Gamma^{[3]}_{tpqrmj} + \sum_{pqr}\delta_{mj}g_{ptqr}\Gamma^{[3]}_{puqrin}  + \sum_{pqr}\delta_{tj}g_{pmqr}\Gamma^{[3]}_{iupnqr} - \sum_{pqr}\delta_{iu}g_{pmqr}\Gamma^{[3]}_{tjpnqr} \\\nonumber
    &\quad - \sum_{pqr}\delta_{iu}g_{ptqr}\Gamma^{[3]}_{pjqrmn} - \sum_{pqr}\delta_{in}g_{pmqr}\Gamma^{[3]}_{tupjqr}  - \sum_{pqr}\delta_{in}g_{ptqr}\Gamma^{[3]}_{puqrmj} - \sum_{pqr}\delta_{mj}g_{npqr}\Gamma^{[3]}_{tuipqr} \\\nonumber
    &\quad - \sum_{pqr}\delta_{mj}g_{upqr}\Gamma^{[3]}_{tpqrin} - \sum_{pqr}\delta_{mj}g_{npqr}\Gamma^{[3]}_{iumpqr}  - \sum_{pqr}\delta_{mj}g_{upqr}\Gamma^{[3]}_{ipqrmn} - \sum_{pqr}\delta_{in}\delta_{mu}g_{ptqr}\Gamma^{[2]}_{pjqr} \bigg)
\end{align}

\begin{align}\label{eq:EE_hE_EE}
    & \left<0\left|\left[\hat{E}_{ij}\hat{E}_{kl}, \left[\sum_{pq}h_{pq}\hat{E}_{pq}, \hat{E}_{tu}\hat{E}_{mn}\right]\right]\right|0\right> \\\nonumber
    & = h_{uk}\Gamma^{[3]}_{ijtlmn} + h_{ui}\Gamma^{[3]}_{tjmnkl} + h_{lt}\Gamma^{[3]}_{ijkumn} + h_{jt}\Gamma^{[3]}_{iumnkl} + h_{nk}\Gamma^{[3]}_{ijtuml} + h_{ni}\Gamma^{[3]}_{tumjkl} \\\nonumber
    &\quad + h_{lm}\Gamma^{[3]}_{ijtukn} + h_{jm}\Gamma^{[3]}_{tuinkl} + \delta_{in}h_{uk}\Gamma^{[2]}_{tlmj} + \delta_{iu}h_{nk}\Gamma^{[2]}_{tjml} + \delta_{kj}h_{ui}\Gamma^{[2]}_{tlmn} + \delta_{kj}h_{lt}\Gamma^{[2]}_{iumn} \\\nonumber
    &\quad + \delta_{kj}h_{ni}\Gamma^{[2]}_{tuml} + \delta_{kj}h_{lm}\Gamma^{[2]}_{intu} + \delta_{kn}h_{ui}\Gamma^{[2]}_{tjml} + \delta_{ku}h_{ni}\Gamma^{[2]}_{tlmj} + \delta_{mj}h_{lt}\Gamma^{[2]}_{inku} + \delta_{ml}h_{jt}\Gamma^{[2]}_{iukn} \\\nonumber
    &\quad + \delta_{mu}h_{lt}\Gamma^{[2]}_{ijkn} + \delta_{mu}h_{jt}\Gamma^{[2]}_{inkl} + \delta_{mu}h_{nk}\Gamma^{[2]}_{ijtl} + \delta_{mu}h_{ni}\Gamma^{[2]}_{tjkl} + \delta_{tj}h_{lm}\Gamma^{[2]}_{iukn} + \delta_{tl}h_{jm}\Gamma^{[2]}_{inku} \\\nonumber
    &\quad + \delta_{kj}\delta_{mu}h_{ni}\Gamma^{[1]}_{tl} + \delta_{kj}\delta_{mu}h_{lt}\Gamma^{[1]}_{in} + \sum_p\delta_{in}h_{up}\Gamma^{[3]}_{tpmjkl} + \sum_p\delta_{iu}h_{np}\Gamma^{[3]}_{tjmpkl} + \sum_p\delta_{kn}h_{up}\Gamma^{[3]}_{ijtpml} \\\nonumber
    &\quad + \sum_p\delta_{ku}h_{np}\Gamma^{[3]}_{ijtlmn} + \sum_p\delta_{mj}h_{pt}\Gamma^{[3]}_{puinkl} + \sum_p\delta_{tj}h_{pm}\Gamma^{[3]}_{iupnkl} + \sum_p\delta_{tl}h_{pm}\Gamma^{[3]}_{ijkupn} \\\nonumber
    &\quad - \sum_p\delta_{in}h_{pt}\Gamma^{[3]}_{pumjkl} - \sum_p\delta_{in}h_{pm}\Gamma^{[3]}_{tupjkl} - \sum_p\delta_{iu}h_{pt}\Gamma^{[3]}_{pjmnkl} - \sum_p\delta_{iu}h_{pm}\Gamma^{[3]}_{tjpnkl} \\\nonumber
    &\quad - \sum_p\delta_{kn}h_{pt}\Gamma^{[3]}_{ijpuml} - \sum_p\delta_{kn}h_{pm}\Gamma^{[3]}_{ijtupl} - \sum_p\delta_{ku}h_{pt}\Gamma^{[3]}_{ijplmn} - \sum_p\delta_{ku}h_{pm}\Gamma^{[3]}_{ijtlpn} \\\nonumber
    &\quad - \sum_p\delta_{mj}h_{up}\Gamma^{[3]}_{tpinkl} - \sum_p\delta_{mj}h_{np}\Gamma^{[3]}_{ijtukp} - \sum_p\delta_{tj}h_{up}\Gamma^{[3]}_{ipmnkl} - \sum_p\delta_{tj}h_{np}\Gamma^{[3]}_{iumpkl} \\\nonumber
    &\quad - \sum_p\delta_{tl}h_{up}\Gamma^{[3]}_{ijkpmn} - \sum_p\delta_{tl}h_{np}\Gamma^{[3]}_{ijkump} + \sum_p\delta_{in}\delta_{kj}h_{up}\Gamma^{[2]}_{tpml} + \sum_p\delta_{iu}\delta_{kj}h_{np}\Gamma^{[2]}_{tlmp} \\\nonumber
    &\quad + \sum_p\delta_{ml}\delta_{kj}h_{pt}\Gamma^{[2]}_{inpu} + \sum_p\delta_{tl}\delta_{kj}h_{pm}\Gamma^{[2]}_{iupn} - \sum_p\delta_{in}\delta_{kj}h_{pt}\Gamma^{[2]}_{puml} - \sum_p\delta_{in}\delta_{kj}h_{pm}\Gamma^{[2]}_{tupl} \\\nonumber
    &\quad - \sum_p\delta_{in}\delta_{ku}h_{pt}\Gamma^{[2]}_{plmj} - \sum_p\delta_{in}\delta_{ku}h_{om}\Gamma^{[2]}_{tlpj} - \sum_p\delta_{in}\delta_{mu}h_{pt}\Gamma^{[2]}_{pjkl} - \sum_p\delta_{iu}\delta_{kj}h_{pm}\Gamma^{[2]}_{tlpn} \\\nonumber
    &\quad - \sum_p\delta_{iu}\delta_{kj}h_{pt}\Gamma^{[2]}_{plmn} - \sum_p\delta_{iu}\delta_{kn}h_{pm}\Gamma^{[2]}_{tjpl} - \sum_p\delta_{iu}\delta_{kn}h_{pt}\Gamma^{[2]}_{pjml} - \sum_p\delta_{kn}\delta_{mu}h_{pt}\Gamma^{[2]}_{ijpl} \\\nonumber
    &\quad - \sum_p\delta_{mj}\delta_{tl}h_{np}\Gamma^{[2]}_{ipku} - \sum_p\delta_{mj}\delta_{tl}h_{up}\Gamma^{[2]}_{inkp} - \sum_p\delta_{ml}\delta_{tj}h_{np}\Gamma^{[2]}_{iukp} - \sum_p\delta_{ml}\delta_{tj}h_{up}\Gamma^{[2]}_{ipkn} \\\nonumber
    &\quad - \sum_p\delta_{ml}\delta_{kj}h_{np}\Gamma^{[2]}_{iptu} - \sum_p\delta_{ml}\delta_{kj}h_{up}\Gamma^{[2]}_{intp} - \sum_p\delta_{tj}\delta_{mu}h_{np}\Gamma^{[2]}_{ipkl} - \sum_p\delta_{tl}\delta_{kj}h_{np}\Gamma^{[2]}_{iump} \\\nonumber
    &\quad - \sum_p\delta_{tl}\delta_{kj}h_{up}\Gamma^{[2]}_{ipmn} - \sum_p\delta_{tl}\delta_{mu}h_{np}\Gamma^{[2]}_{ijkp} - \sum_p\delta_{in}\delta_{kj}\delta_{mu}h_{pt}\Gamma^{[1]}_{pl} - \sum_p\delta_{tl}\delta_{kj}\delta_{mu}h_{np}\Gamma^{[1]}_{ip}
\end{align}

\begin{align}\label{eq:EE_ge_EE}
    & \left<0\left|\left[\hat{E}_{ij}\hat{E}_{kl}, \left[\sum_{pqrs}g_{pqrs}\hat{e}_{pqrs}, \hat{E}_{tu}\hat{E}_{mn}\right]\right]\right|0\right> \\\nonumber
    & = 2\left( \mathcal{A} + \mathcal{B} + \mathcal{C} + \mathcal{D} + \mathcal{E} + \mathcal{F} + \mathcal{G} + \mathcal{H} \right)
\end{align}

\begin{align}\label{eq:EE_ge_EE:A}
    \mathcal{A} = g_{niuk}\Gamma^{[2]}_{tlmj} + g_{nkui}\Gamma^{[2]}_{tjml} + g_{ltjm}\Gamma^{[2]}_{inku} + g_{jtlm}\Gamma^{[2]}_{iukn}
\end{align}

\begin{align}\label{eq:EE_ge_EE:B}
    \mathcal{B} = & \sum_pg_{lmjp}\Gamma^{[3]}_{tuipkn} + \sum_pg_{jmlp}\Gamma^{[3]}_{tuinkp} + \sum_pg_{nkup}\Gamma^{[3]}_{tpijml} + \sum_pg_{npuk}\Gamma^{[3]}_{tlijmp} \\\nonumber
    + & \sum_pg_{nipk}\Gamma^{[3]}_{tumjpl} + \sum_pg_{niup}\Gamma^{[3]}_{tpmjpl} + \sum_pg_{npui}\Gamma^{[3]}_{tjmpkl} + \sum_pg_{nkpi}\Gamma^{[3]}_{tumlpj} \\\nonumber
    + & \sum_pg_{ltpm}\Gamma^{[3]}_{ijkupn} + \sum_pg_{ltjp}\Gamma^{[3]}_{ipkumn} + \sum_pg_{ptlm}\Gamma^{[3]}_{ijpukn} + \sum_pg_{jtlp}\Gamma^{[3]}_{iukpmn} \\\nonumber
    + & \sum_pg_{jtpm}\Gamma^{[3]}_{iupnkl} + \sum_pg_{ptjm}\Gamma^{[3]}_{puinkl} + \sum_pg_{uipk}\Gamma^{[3]}_{tjplmn} + \sum_pg_{ukpi}\Gamma^{[3]}_{tlpjmn} \\\nonumber
    + & \sum_p\delta_{kj}g_{niup}\Gamma^{[2]}_{tpml} + \sum_p\delta_{kj}g_{npui}\Gamma^{[2]}_{tlmp} + \sum_p\delta_{kj}g_{ltpm}\Gamma^{[2]}_{iupn} + \sum_p\delta_{kj}g_{ptlm}\Gamma^{[2]}_{inpu} \\\nonumber
    + & \sum_p\delta_{mu}g_{nipk}\Gamma^{[2]}_{tjpl} + \sum_p\delta_{mu}g_{nkpi}\Gamma^{[2]}_{tlpj} + \sum_p\delta_{mu}g_{ltjp}\Gamma^{[2]}_{ipkn} + \sum_p\delta_{mu}g_{jtlp}\Gamma^{[2]}_{inkp}
\end{align}

\begin{align}\label{eq:EE_ge_EE:C}
    \mathcal{C} = & \sum_{pq}g_{lmpq}\Gamma^{[4]}_{tuijknpq} + \sum_{pq}g_{pmlq}\Gamma^{[4]}_{tuijpnkq} + \sum_{pq}g_{nkpq}\Gamma^{[4]}_{tuijmlpq} + \sum_{pq}g_{npqk}\Gamma^{[4]}_{tuijmpql} \\\nonumber
    + & \sum_{pq}g_{jmpq}\Gamma^{[4]}_{tuinpqkl} + \sum_{pq}g_{pmjq}\Gamma^{[4]}_{tupniqkl} + \sum_{pq}g_{nipq}\Gamma^{[4]}_{tumjpqkl} + \sum_{pq}g_{npqi}\Gamma^{[4]}_{tumpqjkl} \\\nonumber
    + & \sum_{pq}g_{ltpq}\Gamma^{[4]}_{ijkupqmn} + \sum_{pq}g_{ptlq}\Gamma^{[4]}_{ijpukqmn} + \sum_{pq}g_{ukpq}\Gamma^{[4]}_{ijtlpqmn} + \sum_{pq}g_{upqk}\Gamma^{[4]}_{ijtpqlmn} \\\nonumber
    + & \sum_{pq}g_{jtpq}\Gamma^{[4]}_{iupqklmn} + \sum_{pq}g_{ptjq}\Gamma^{[4]}_{puiqklmn} + \sum_{pq}g_{uipq}\Gamma^{[4]}_{tjpqklmn} + \sum_{pq}g_{upqi}\Gamma^{[4]}_{tpqjklmn} \\\nonumber
    - & \sum_{pq}g_{nplq}\Gamma^{[4]}_{tuijmpkq} - \sum_{pq}g_{pmqk}\Gamma^{[4]}_{tuijpnql} - \sum_{pq}g_{npjq}\Gamma^{[4]}_{tumpiqkl} - \sum_{pq}g_{pmqi}\Gamma^{[4]}_{tupnqjkl} \\\nonumber
    - & \sum_{pq}g_{ptqk}\Gamma^{[4]}_{ijpuqlmn} - \sum_{pq}g_{uplq}\Gamma^{[4]}_{ijtpkqmn} - \sum_{pq}g_{ptqi}\Gamma^{[4]}_{puqjklmn} - \sum_{pq}g_{upjq}\Gamma^{[4]}_{tpiqklmn}
\end{align}

\begin{align}\label{eq:EE_ge_EE:D}
    \mathcal{D} = & \sum_{pq}\delta_{in}g_{upqk}\Gamma^{[3]}_{tpqlmj} + \sum_{pq}\delta_{in}g_{ukpq}\Gamma^{[3]}_{tlpqmj} + \sum_{pq}\delta_{iu}g_{nkpq}\Gamma^{[3]}_{tjmlpq} + \sum_{pq}\delta_{iu}g_{npqk}\Gamma^{[3]}_{tjmpql}\\\nonumber
    & + \sum_{pq}\delta_{kj}g_{lmpq}\Gamma^{[3]}_{tuinpq} + \sum_{pq}\delta_{kj}g_{pmlq}\Gamma^{[3]}_{tuiqpn} + \sum_{pq}\delta_{kj}g_{nipq}\Gamma^{[3]}_{tumlpq} + \sum_{pq}\delta_{kj}g_{npqi}\Gamma^{[3]}_{tumpql}\\\nonumber
    & + \sum_{pq}\delta_{kj}g_{ltpq}\Gamma^{[3]}_{iupqmn} + \sum_{pq}\delta_{kj}g_{ptlq}\Gamma^{[3]}_{iqpumn} + \sum_{pq}\delta_{kj}g_{uipq}\Gamma^{[3]}_{tlpqmn} + \sum_{pq}\delta_{kj}g_{upqi}\Gamma^{[3]}_{tlpqmn}\\\nonumber
    & + \sum_{pq}\delta_{kn}g_{upqi}\Gamma^{[3]}_{tpqjml} + \sum_{pq}\delta_{kn}g_{uipq}\Gamma^{[3]}_{tjpqml} + \sum_{pq}\delta_{ku}g_{nipq}\Gamma^{[3]}_{tlmjpq} + \sum_{pq}\delta_{ku}g_{npqi}\Gamma^{[3]}_{tlmpqj}\\\nonumber
    & + \sum_{pq}\delta_{mj}g_{ltpq}\Gamma^{[3]}_{inkupq} + \sum_{pq}\delta_{mj}g_{ptlq}\Gamma^{[3]}_{inpukq} + \sum_{pq}\delta_{mu}g_{nkpq}\Gamma^{[3]}_{tlijpq} + \sum_{pq}\delta_{mu}g_{npqk}\Gamma^{[3]}_{tpijql}\\\nonumber
    & + \sum_{pq}\delta_{mu}g_{nipq}\Gamma^{[3]}_{tjpqkl} + \sum_{pq}\delta_{mu}g_{npqi}\Gamma^{[3]}_{tpqjkl} + \sum_{pq}\delta_{mu}g_{ltpq}\Gamma^{[3]}_{ijknpq} + \sum_{pq}\delta_{mu}g_{ptlq}\Gamma^{[3]}_{ijpnkq}\\\nonumber
    & + \sum_{pq}\delta_{mu}g_{jtpq}\Gamma^{[3]}_{inpqkl} + \sum_{pq}\delta_{mu}g_{ptjq}\Gamma^{[3]}_{pniqkl} + \sum_{pq}\delta_{ml}g_{jtpq}\Gamma^{[3]}_{iupqkn} + \sum_{pq}\delta_{ml}g_{ptjq}\Gamma^{[3]}_{puiqkn}\\\nonumber
    & + \sum_{pq}\delta_{tl}g_{jmpq}\Gamma^{[3]}_{inkupq} + \sum_{pq}\delta_{tl}g_{pmjq}\Gamma^{[3]}_{iqkupn} + \sum_{pq}\delta_{tj}g_{lmpq}\Gamma^{[3]}_{iuknpq} + \sum_{pq}\delta_{tj}g_{pmlq}\Gamma^{[3]}_{iukqpn}\\\nonumber
    & - \sum_{pq}\delta_{in}g_{ptqm}\Gamma^{[3]}_{puqjkl} - \sum_{pq}\delta_{in}g_{ptqk}\Gamma^{[3]}_{puqlmj} - \sum_{pq}\delta_{in}g_{pmqk}\Gamma^{[3]}_{tupjql} - \sum_{pq}\delta_{iu}g_{pmqk}\Gamma^{[3]}_{tjpnql}\\\nonumber
    & - \sum_{pq}\delta_{iu}g_{ptqk}\Gamma^{[3]}_{pjqlmn} - \sum_{pq}\delta_{iu}g_{ptqm}\Gamma^{[3]}_{pjqnkl} - \sum_{pq}\delta_{kj}g_{nplq}\Gamma^{[3]}_{tuiqmp} - \sum_{pq}\delta_{kj}g_{pmqi}\Gamma^{[3]}_{tupnql}\\\nonumber
    & - \sum_{pq}\delta_{kj}g_{uplq}\Gamma^{[3]}_{iqtpmn} - \sum_{pq}\delta_{kj}g_{ptqi}\Gamma^{[3]}_{puqlmn} - \sum_{pq}\delta_{kn}g_{pmqi}\Gamma^{[3]}_{tuplqj} - \sum_{pq}\delta_{kn}g_{ptqi}\Gamma^{[3]}_{puqjml}\\\nonumber
    & - \sum_{pq}\delta_{kn}g_{ptqm}\Gamma^{[3]}_{puqlij} - \sum_{pq}\delta_{ku}g_{pmqi}\Gamma^{[3]}_{tlpnqj} - \sum_{pq}\delta_{ku}g_{ptqi}\Gamma^{[3]}_{plqjmn} - \sum_{pq}\delta_{ku}g_{ptqm}\Gamma^{[3]}_{ijplqn}\\\nonumber
    & - \sum_{pq}\delta_{mj}g_{nplq}\Gamma^{[3]}_{tuipkq} - \sum_{pq}\delta_{mj}g_{uplq}\Gamma^{[3]}_{intpkq} - \sum_{pq}\delta_{mj}g_{npuq}\Gamma^{[3]}_{tqipkl} - \sum_{pq}\delta_{mu}g_{nplq}\Gamma^{[3]}_{tpijkq}\\\nonumber
    & - \sum_{pq}\delta_{mu}g_{npjq}\Gamma^{[3]}_{tpiqkl} - \sum_{pq}\delta_{mu}g_{ptqk}\Gamma^{[3]}_{ijpnql} - \sum_{pq}\delta_{mu}g_{ptqi}\Gamma^{[3]}_{pnqjkl} - \sum_{pq}\delta_{ml}g_{upjq}\Gamma^{[3]}_{tpiqkn}\\\nonumber
    & - \sum_{pq}\delta_{ml}g_{npuq}\Gamma^{[3]}_{tqijkp} - \sum_{pq}\delta_{ml}g_{npjq}\Gamma^{[3]}_{tuiqkp} - \sum_{pq}\delta_{tj}g_{uplq}\Gamma^{[3]}_{ipkqmn} - \sum_{pq}\delta_{tj}g_{npuq}\Gamma^{[3]}_{iqklmp}\\\nonumber
    & - \sum_{pq}\delta_{tl}g_{npuq}\Gamma^{[3]}_{ijkqmp} - \sum_{pq}\delta_{tl}g_{npjq}\Gamma^{[3]}_{iqkump} - \sum_{pq}\delta_{tl}g_{upjq}\Gamma^{[3]}_{iqkpmn}
\end{align}

\begin{align}\label{eq:EE_ge_EE:E}
    \mathcal{E} = & \sum_{pq}\delta_{kj}\delta_{mu}g_{nipq}\Gamma^{[2]}_{tlpq} + \sum_{pq}\delta_{kj}\delta_{mu}g_{npqi}\Gamma^{[2]}_{tpql} + \sum_{pq}\delta_{kj}\delta_{mu}g_{ltpq}\Gamma^{[2]}_{inpq} \\\nonumber
    &\quad + \sum_{pq}\delta_{kj}\delta_{mu}g_{ptlq}\Gamma^{[2]}_{iqpn} + \sum_{pq}\delta_{kj}\delta_{tl}g_{pmuq}\Gamma^{[2]}_{iqpn} - \sum_{pq}\delta_{in}\delta_{kj}g_{ptqm}\Gamma^{[2]}_{puql} \\\nonumber
    &\quad - \sum_{pq}\delta_{in}\delta_{ku}g_{ptqm}\Gamma^{[2]}_{plqj} - \sum_{pq}\delta_{in}\delta_{mu}g_{ptqk}\Gamma^{[2]}_{pjql} - \sum_{pq}\delta_{iu}\delta_{kj}g_{ptqm}\Gamma^{[2]}_{plqn} \\\nonumber
    &\quad - \sum_{pq}\delta_{iu}\delta_{kn}g_{ptqm}\Gamma^{[2]}_{pjql} - \sum_{pq}\delta_{kj}\delta_{mu}g_{nplq}\Gamma^{[2]}_{tpiq} - \sum_{pq}\delta_{kj}\delta_{mu}g_{ptqi}\Gamma^{[2]}_{pnql} \\\nonumber
    &\quad - \sum_{pq}\delta_{kj}\delta_{ml}g_{npuq}\Gamma^{[2]}_{tqip} - \sum_{pq}\delta_{kj}\delta_{tl}g_{upqm}\Gamma^{[2]}_{ipqn} - \sum_{pq}\delta_{kj}\delta_{tl}g_{npuq}\Gamma^{[2]}_{iqmp} \\\nonumber
    &\quad - \sum_{pq}\delta_{kn}\delta_{mu}g_{ptqi}\Gamma^{[2]}_{plqj} - \sum_{pq}\delta_{mj}\delta_{tl}g_{npuq}\Gamma^{[2]}_{ipkq} - \sum_{pq}\delta_{mu}\delta_{tj}g_{nplq}\Gamma^{[2]}_{ipkq} \\\nonumber
    &\quad - \sum_{pq}\delta_{mu}\delta_{tl}g_{npjq}\Gamma^{[2]}_{iqkp} - \sum_{pq}\delta_{ml}\delta_{tj}g_{npuq}\Gamma^{[2]}_{iqkp}
\end{align}

\begin{align}\label{eq:EE_ge_EE:F}
    \mathcal{F} = & \sum_{pqr}\delta_{in}g_{upqr}\Gamma^{[4]}_{tpqrmjkl} + \sum_{pqr}\delta_{iu}g_{npqr}\Gamma^{[4]}_{tjklmpqr} + \sum_{pqr}\delta_{kn}g_{upqr}\Gamma^{[4]}_{tpqrijml} \\\nonumber
    & + \sum_{pqr}\delta_{ku}g_{npqr}\Gamma^{[4]}_{ijtlmpqr} + \sum_{pqr}\delta_{mj}g_{ptqr}\Gamma^{[4]}_{puqrinkl} + \sum_{pqr}\delta_{ml}g_{ptqr}\Gamma^{[4]}_{puqrijkn} \\\nonumber
    & + \sum_{pqr}\delta_{tj}g_{pmqr}\Gamma^{[4]}_{iuklpnqr} + \sum_{pqr}\delta_{tl}g_{pmqr}\Gamma^{[4]}_{ijkupnqr} - \sum_{pqr}\delta_{in}g_{ptqr}\Gamma^{[4]}_{puqrmjkl} \\\nonumber
    & - \sum_{pqr}\delta_{in}g_{pmqr}\Gamma^{[4]}_{tupjqrkl} - \sum_{pqr}\delta_{iu}g_{pmqr}\Gamma^{[4]}_{tjklpnqr} - \sum_{pqr}\delta_{iu}g_{ptqr}\Gamma^{[4]}_{pjqrklmn} \\\nonumber
    & - \sum_{pqr}\delta_{kn}g_{ptqr}\Gamma^{[4]}_{puqrijml} - \sum_{pqr}\delta_{kn}g_{pmqr}\Gamma^{[4]}_{tuijplqr} - \sum_{pqr}\delta_{ku}g_{pmqr}\Gamma^{[4]}_{ijtlpnqr} \\\nonumber
    & - \sum_{pqr}\delta_{ku}g_{ptqr}\Gamma^{[4]}_{ijplqrmn} - \sum_{pqr}\delta_{mj}g_{upqr}\Gamma^{[4]}_{tpqrinkl} - \sum_{pqr}\delta_{mj}g_{npqr}\Gamma^{[4]}_{tuipqrkl} \\\nonumber
    & - \sum_{pqr}\delta_{ml}g_{upqr}\Gamma^{[4]}_{tpqrijkn} - \sum_{pqr}\delta_{ml}g_{npqr}\Gamma^{[4]}_{tuijkpqr} - \sum_{pqr}\delta_{tj}g_{npqr}\Gamma^{[4]}_{iuklmpqr} \\\nonumber
    & - \sum_{pqr}\delta_{tj}g_{upqr}\Gamma^{[4]}_{ipqrklmn} - \sum_{pqr}\delta_{tl}g_{npqr}\Gamma^{[4]}_{ijkumpqr} - \sum_{pqr}\delta_{tl}g_{upqr}\Gamma^{[4]}_{ijkpqrmn}
\end{align}

\begin{align}\label{eq:EE_ge_EE:G}
    \mathcal{G} = & \sum_{pqr}\delta_{in}\delta_{kj}g_{upqr}\Gamma^{[3]}_{tpqrml} + \sum_{pqr}\delta_{iu}\delta_{kj}g_{npqr}\Gamma^{[3]}_{tlmpqr} + \sum_{pqr}\delta_{ml}\delta_{kj}g_{ptqr}\Gamma^{[3]}_{puqrin} \\\nonumber
    & + \sum_{pqr}\delta_{tl}\delta_{kj}g_{pmqr}\Gamma^{[3]}_{iupnqr} - \sum_{pqr}\delta_{in}\delta_{kj}g_{ptqr}\Gamma^{[3]}_{puqrml} - \sum_{pqr}\delta_{in}\delta_{kj}g_{pmqr}\Gamma^{[3]}_{tuplqr} \\\nonumber
    & - \sum_{pqr}\delta_{in}\delta_{ku}g_{ptqr}\Gamma^{[3]}_{plqrmj} - \sum_{pqr}\delta_{in}\delta_{ku}g_{pmqr}\Gamma^{[3]}_{tlpjqr} - \sum_{pqr}\delta_{in}\delta_{mu}g_{ptqr}\Gamma^{[3]}_{pjqrkl} \\\nonumber
    & - \sum_{pqr}\delta_{iu}\delta_{kn}g_{ptqr}\Gamma^{[3]}_{pjqrml} - \sum_{pqr}\delta_{iu}\delta_{kn}g_{pmqr}\Gamma^{[3]}_{tjplqr} - \sum_{pqr}\delta_{iu}\delta_{kj}g_{ptqr}\Gamma^{[3]}_{plqrmn} \\\nonumber
    & - \sum_{pqr}\delta_{iu}\delta_{kj}g_{pmqr}\Gamma^{[3]}_{tlpnqr} - \sum_{pqr}\delta_{kn}\delta_{mu}g_{ptqr}\Gamma^{[3]}_{plqrij} - \sum_{pqr}\delta_{mj}\delta_{tl}g_{npqr}\Gamma^{[3]}_{ipkuqr} \\\nonumber
    & - \sum_{pqr}\delta_{mj}\delta_{tl}g_{upqr}\Gamma^{[3]}_{inkpqr} - \sum_{pqr}\delta_{ml}\delta_{kj}g_{upqr}\Gamma^{[3]}_{tpqrin} - \sum_{pqr}\delta_{ml}\delta_{kj}g_{npqr}\Gamma^{[3]}_{tuinqr} \\\nonumber
    & - \sum_{pqr}\delta_{ml}\delta_{tj}g_{npqr}\Gamma^{[3]}_{iukpqr} - \sum_{pqr}\delta_{ml}\delta_{tj}g_{upqr}\Gamma^{[3]}_{ipqrkn} - \sum_{pqr}\delta_{tj}\delta_{mu}g_{npqr}\Gamma^{[3]}_{ipklqr} \\\nonumber
    & - \sum_{pqr}\delta_{tl}\delta_{kj}g_{npqr}\Gamma^{[3]}_{iumpqr} - \sum_{pqr}\delta_{tl}\delta_{kj}g_{upqr}\Gamma^{[3]}_{ipqrmn} - \sum_{pqr}\delta_{tl}\delta_{mu}g_{npqr}\Gamma^{[3]}_{ijkpqr}
\end{align}

\begin{align}\label{eq:EE_ge_EE:H}
    \mathcal{H} = - \sum_{pqr}\delta_{in}\delta_{kj}\delta_{mu}g_{ptqr}\Gamma^{[2]}_{plqr} - \sum_{pqr}\delta_{tl}\delta_{kj}\delta_{mu}g_{npqr}\Gamma^{[2]}_{ipqr}
\end{align}

\begin{equation}\label{eq:E_E}
    \left[\hat{E}_{mn}, \hat{E}_{pq}\right] = \delta_{pn}\Gamma^{[1]}_{mq} - \delta_{mq}\Gamma^{[1]}_{pn}
\end{equation}

\begin{align}\label{eq:E_EE}
    &\left<0\left|\left[\hat{E}_{mn},\hat{E}_{pq}\hat{E}_{rs}\right]\right|0\right> \\\nonumber
    & = \delta_{pn}\Gamma^{[2]}_{mqrs} + \delta_{pn}\delta_{qr}\Gamma^{[1]}_{ms} - \delta_{mq}\Gamma^{[2]}_{pnrs} \\\nonumber
    &\quad + \delta_{rn}\Gamma^{[2]}_{pqms} - \delta_{ms}\Gamma^{[2]}_{pqrn} - \delta_{ms}\delta_{qr}\Gamma^{[1]}_{pn}
\end{align}

\begin{align}\label{eq:EE_EE}
    &\left<0\left|\left[\hat{E}_{tu}\hat{E}_{mn}, \hat{E}_{pq}\hat{E}_{rs}\right]\right|0\right> \\\nonumber
    & = - {}^-P_{qn}^{mp}\delta_{mq} \bigg( \Gamma^{[3]}_{tupnrs} + \delta_{nr} \left( \Gamma^{[2]}_{tups} + \delta_{up} \Gamma^{[1]}_{ts} \right) + \delta_{up}\Gamma^{[2]}_{tnrs} + \delta_{ur}\Gamma^{[2]}_{tspn} \bigg) \\\nonumber
    &\quad - {}^-P_{sn}^{mr}\delta_{ms} \bigg( \Gamma^{[3]}_{tupqrn} + \delta_{qr} \left( \Gamma^{[2]}_{tupn} + \delta_{up} \Gamma^{[1]}_{tn} \right) + \delta_{up}\Gamma^{[2]}_{tqrn} + \delta_{ur}\Gamma^{[2]}_{tnpq} \bigg) \\\nonumber
    &\quad - {}^-P_{qu}^{tp}\delta_{tq} \bigg( \Gamma^{[3]}_{pursmn} + \delta_{sm} \left( \Gamma^{[2]}_{purn} + \delta_{ur} \Gamma^{[1]}_{pn} \right) + \delta_{ur}\Gamma^{[2]}_{psmn} + \delta_{um}\Gamma^{[2]}_{pnrs} \bigg) \\\nonumber
    &\quad - {}^-P_{su}^{tr}\delta_{ts} \bigg( \Gamma^{[3]}_{pqrumn} + \delta_{um} \left( \Gamma^{[2]}_{pqrn} + \delta_{qr} \Gamma^{[1]}_{pn} \right) + \delta_{qr}\Gamma^{[2]}_{pumn} + \delta_{qm}\Gamma^{[2]}_{pnru} \bigg)
\end{align}

\section{Geometries}\label{sec:geom}

Benzene:
\begin{verbatim}
C   -1.210   0.698   0.004
C   -1.210  -0.698   0.000
C    0.000  -1.397  -0.003
C    1.210  -0.698  -0.003
C    1.210   0.698   0.000
C    0.000   1.397  -0.003
H   -2.164   1.249   0.007
H   -2.164  -1.249   0.007
H    0.000  -2.500  -0.005
H    2.164  -1.249  -0.006
H    2.164   1.249   0.000
H    0.000   2.500   0.000
\end{verbatim}
\clearpage
$R$-methyloxirane:
\begin{verbatim}
O   0.8171  -0.7825  -0.2447
C  -1.5018   0.1019  -0.1473
H  -1.3989   0.3323  -1.2066
H  -2.0652  -0.8262  -0.0524
H  -2.0715   0.8983   0.3329
C  -0.1488  -0.0393   0.4879
H  -0.1505  -0.2633   1.5506
C   1.0387   0.6105  -0.0580
H   0.9518   1.2157  -0.9531
H   1.8684   0.8649   0.5908
\end{verbatim}

Water:
\begin{verbatim}
O   0.0  0.0           0.1035174918
H   0.0  0.7955612117 -0.4640237459
H   0.0 -0.7955612117 -0.4640237459
\end{verbatim}

\section{Code examples}
\label{sec:code_examples}

\begin{figure}[H]
    \noindent\makebox[\linewidth]{\rule{\textwidth}{0.6pt}}
    \vspace{-.5cm}
    \begin{algorithmic}
        \STATE \texttt{\textcolor{OliveGreen}{from} \textcolor{blue}{dmdm.interface} \textcolor{OliveGreen}{import} DMDM}
        \STATE 
        \STATE \texttt{dmdm = DMDM(}
        \STATE \quad \qquad \texttt{int1e, int2e,}
        \STATE \quad \qquad \texttt{num\_inact\_orb, num\_act\_orb, num\_virt\_orb, num\_elec,}
        \STATE \quad \qquad \texttt{rdm1, rdm2, rdm3, rdm4}
        \STATE )
        \STATE \texttt{excitation\_energies = dmdm.get\_excitation\_energies()}
        \STATE \texttt{oscillator\_strengths = dmdm.get\_oscillator\_strength(elec\_dip\_int)}
    \end{algorithmic}
    \vspace{-.4cm}
    \noindent\makebox[\linewidth]{\rule{\textwidth}{0.6pt}}
\caption{Excitation energies and oscillator strengths calculation.
\texttt{int1e}  
are the one-electron integrals from the Hamiltonian in MO. 
\texttt{int2e} 
are the two-electron integrals from the Hamiltonian in MO; 
\texttt{num\_inact\_orb} is the number of inactive spatial orbitals; \texttt{num\_act\_orb} is the number of active spatial orbitals; \texttt{num\_virt\_orb} is the number of virtual spatial orbitals;\texttt{num\_elec} is the number of electrons. \texttt{rdm1}, \texttt{rdm2}, \texttt{rdm3} and \texttt{rdm4} are the n-RDMs in only the active space. \texttt{elec\_dip\_int} are the electric dipole integrals in MO ($x$, $y$, $z$ components).}
\end{figure}

\begin{figure}[H]
    \noindent\makebox[\linewidth]{\rule{\textwidth}{0.6pt}}
    \vspace{-.5cm}
    \begin{algorithmic}
        \STATE \texttt{\textcolor{OliveGreen}{from} \textcolor{blue}{dmdm.interface} \textcolor{OliveGreen}{import} DMDM}
        \STATE 
        \STATE \texttt{dmdm = DMDM(}
        \STATE \quad \qquad \texttt{int1e, int2e,}
        \STATE \quad \qquad \texttt{num\_inact\_orb, num\_act\_orb, num\_virt\_orb, num\_elec,}
        \STATE \quad \qquad \texttt{rdm1, rdm2, rdm3, rdm4}
        \STATE )
        \STATE \texttt{excitation\_energies = dmdm.get\_excitation\_energies()}
        \STATE \texttt{rotational\_strengths = dmdm.get\_rotational\_strength(elec\_dip\_int, mag\_dip\_int)}
    \end{algorithmic}
    \vspace{-.4cm}
    \noindent\makebox[\linewidth]{\rule{\textwidth}{0.6pt}}
\caption{Excitation energies and rotational strengths calculation.
\texttt{int1e} are the one-electron integrals from the Hamiltonian in MO. 
\texttt{int2e} are the two-electron integrals from the Hamiltonian in MO. 
\texttt{num\_inact\_orb} is the number of inactive spatial orbitals. \texttt{num\_act\_orb} is the number of active spatial orbitals. \texttt{num\_virt\_orb} is the number of virtual spatial orbitals. \texttt{num\_elec} is the number of electrons. \texttt{rdm1}, \texttt{rdm2}, \texttt{rdm3}, and \texttt{rdm4} are the n-RDMs in only the active space. \texttt{elec\_dip\_int} are the electric dipole integrals in MO ($x$, $y$, $z$); \texttt{mag\_dip\_int} are the magnetic dipole integrals in MO ($x$, $y$, $z$)}
\end{figure}

\section{Tables and figures}\label{sec:TabFig}

\begin{figure}[H]
    \centering
    \includegraphics[width=.7\textwidth]{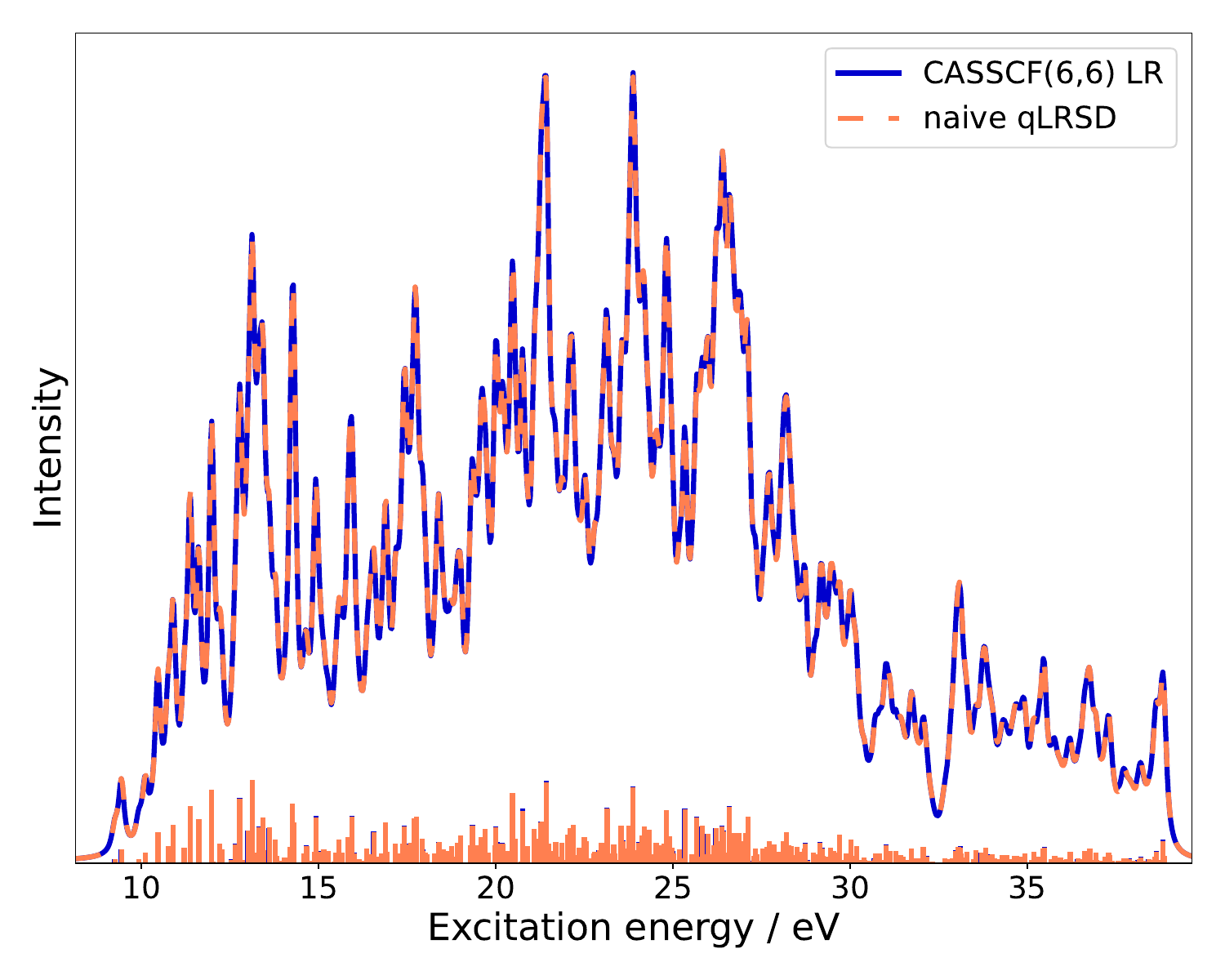}
    \caption{$R$-methyloxirane. Absorption spectra of the first 399 excitation energies using CASSCF(6,6) LR (blue), and qLRSD (dashed orange) based on the oo-UCCSD(6,6) wave function using the cc-pVTZ basis set.}
    \label{fig:absorption_methyloxirane}
\end{figure}

\begin{figure}[H]
    \centering
    \includegraphics[width=.7\textwidth]{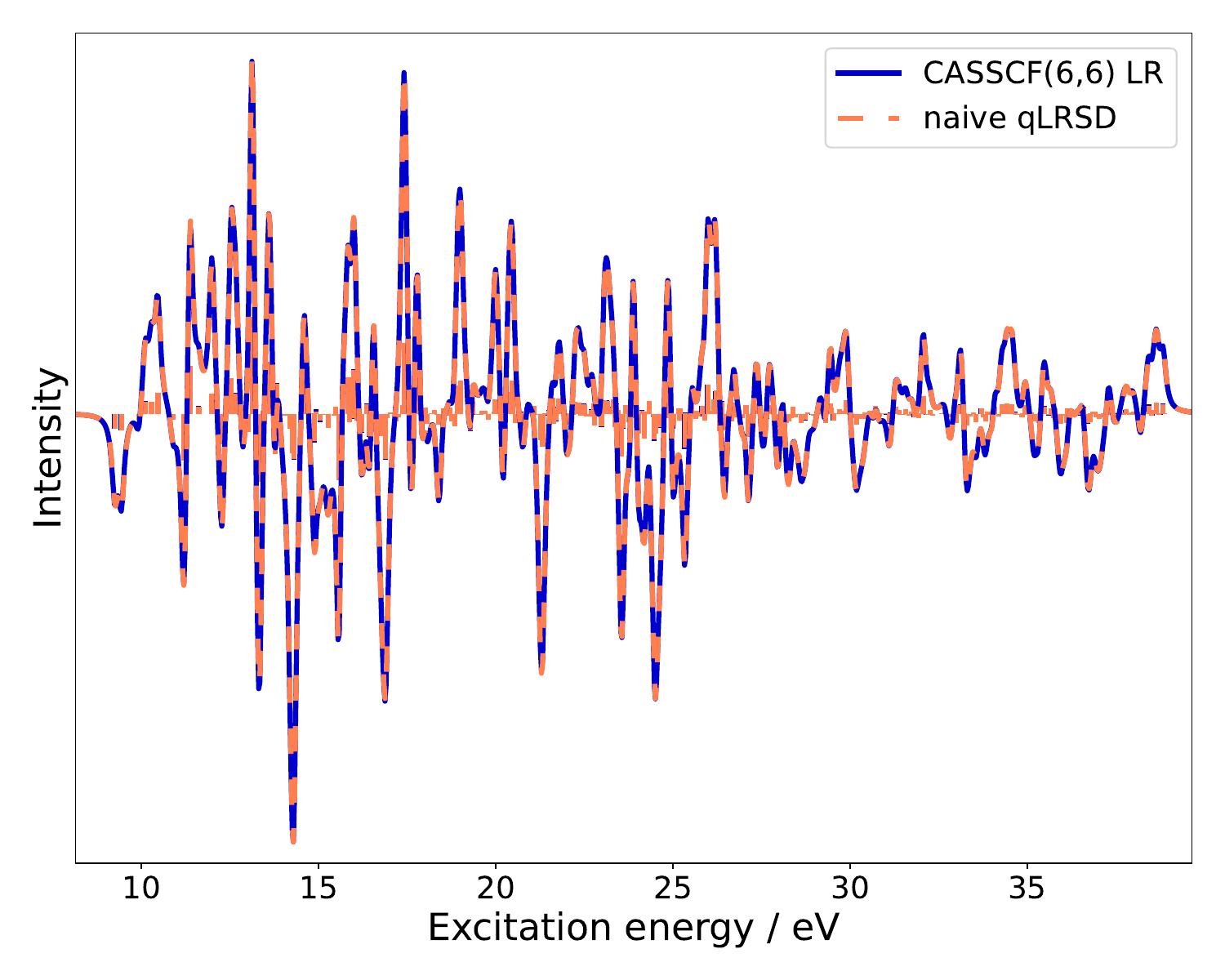}
    \caption{$R$-methyloxirane. ECD spectra of the first 399 excitation energies using CASSCF(6,6) LR (blue), and qLRSD (dashed orange) based on the oo-UCCSD(6,6) wave function using the cc-pVTZ basis set.}
    \label{fig:ecd_full}
\end{figure}

\clearpage

\begin{sidewaystable}
\begin{table}[H]
    \centering
    \begin{tabular}{cc|cccccc|cccccc}
     & & \multicolumn{6}{c|}{Variance} & \multicolumn{6}{c}{Standard deviation} \\
    \multicolumn{2}{c|}{Ideal} & \multicolumn{2}{c}{10k shots} & \multicolumn{2}{c}{50k shots} & \multicolumn{2}{c|}{100k shots} & \multicolumn{2}{c}{10k shots} & \multicolumn{2}{c}{50k shots} & \multicolumn{2}{c}{100k shots} \\
    Exc. & Osc. & Exc. & Osc. & Exc. & Osc. & Exc. & Osc. & Exc. & Osc. & Exc. & Osc. & Exc. & Osc. \\\hline\hline
8.5433 & 0.0319 & 0.1605 & 0.0000 & 0.0084 & 0.0000 & 0.0008 & 0.0000 & 0.4007 & 0.0034 & 0.0915 & 0.0008 & 0.0287 & 0.0003 \\
10.4364 & 0.0000 & 0.0118 & 0.0000 & 0.0028 & 0.0000 & 0.0008 & 0.0000 & 0.1086 & 0.0018 & 0.0529 & 0.0002 & 0.0290 & 0.0000 \\
11.1083 & 0.1133 & 0.0043 & 0.0000 & 0.0006 & 0.0000 & 0.0002 & 0.0000 & 0.0657 & 0.0008 & 0.0248 & 0.0005 & 0.0134 & 0.0003 \\
12.8695 & 0.0976 & 0.0054 & 0.0000 & 0.0009 & 0.0000 & 0.0002 & 0.0000 & 0.0734 & 0.0036 & 0.0307 & 0.0017 & 0.0147 & 0.0012 \\
14.8841 & 0.2517 & 0.0429 & 0.0116 & 0.0050 & 0.0000 & 0.0019 & 0.0000 & 0.2072 & 0.1076 & 0.0708 & 0.0020 & 0.0430 & 0.0018 \\\hline\hline
546.6519 & 0.0427 & 0.1795 & 0.0001 & 0.0329 & 0.0000 & 0.0099 & 0.0000 & 0.4237 & 0.0106 & 0.1813 & 0.0027 & 0.0994 & 0.0016 \\
547.0419 & 0.0735 & 0.1260 & 0.0002 & 0.0523 & 0.0000 & 0.0062 & 0.0000 & 0.3549 & 0.0124 & 0.2287 & 0.0034 & 0.0787 & 0.0016 \\
560.3159 & 0.0303 & 0.0094 & 0.0000 & 0.0043 & 0.0000 & 0.0013 & 0.0000 & 0.0970 & 0.0017 & 0.0658 & 0.0008 & 0.0362 & 0.0006 \\
561.5096 & 0.0014 & 0.0282 & 0.0000 & 0.0064 & 0.0000 & 0.0006 & 0.0000 & 0.1678 & 0.0005 & 0.0800 & 0.0002 & 0.0248 & 0.0001 \\
561.7025 & 0.0779 & 0.0040 & 0.0000 & 0.0011 & 0.0000 & 0.0003 & 0.0000 & 0.0632 & 0.0001 & 0.0336 & 0.0000 & 0.0171 & 0.0000 \\
563.5124 & 0.0094 & 0.0067 & 0.0000 & 0.0020 & 0.0000 & 0.0005 & 0.0000 & 0.0819 & 0.0005 & 0.0444 & 0.0002 & 0.0226 & 0.0002 \\
567.0508 & 0.0236 & 0.0069 & 0.0000 & 0.0019 & 0.0000 & 0.0004 & 0.0000 & 0.0833 & 0.0009 & 0.0435 & 0.0004 & 0.0200 & 0.0001 \\
569.4280 & 0.0001 & 0.0079 & 0.0000 & 0.0031 & 0.0000 & 0.0008 & 0.0000 & 0.0891 & 0.0001 & 0.0560 & 0.0000 & 0.0276 & 0.0000 \\
570.4970 & 0.0000 & 0.0066 & 0.0000 & 0.0018 & 0.0000 & 0.0005 & 0.0000 & 0.0813 & 0.0000 & 0.0425 & 0.0000 & 0.0217 & 0.0000 \\
573.9153 & 0.0000 & 0.0062 & 0.0000 & 0.0017 & 0.0000 & 0.0004 & 0.0000 & 0.0789 & 0.0000 & 0.0415 & 0.0000 & 0.0211 & 0.0000 \\
574.3891 & 0.0021 & 0.0052 & 0.0000 & 0.0014 & 0.0000 & 0.0004 & 0.0000 & 0.0721 & 0.0003 & 0.0375 & 0.0000 & 0.0191 & 0.0000 \\
580.8528 & 0.0003 & 2.5928 & 0.0000 & 0.0222 & 0.0000 & 0.0005 & 0.0000 & 1.6102 & 0.0068 & 0.1488 & 0.0040 & 0.0228 & 0.0000 \\
586.0916 & 0.0135 & 6.4146 & 0.0000 & 3.5990 & 0.0000 & 0.2938 & 0.0000 & 2.5327 & 0.0063 & 1.8971 & 0.0038 & 0.5420 & 0.0032 \\
586.3814 & 0.0267 & 0.5565 & 0.0001 & 0.0901 & 0.0001 & 0.0259 & 0.0001 & 0.7460 & 0.0106 & 0.3002 & 0.0081 & 0.1610 & 0.0084 \\
587.7218 & 0.0007 & 0.6310 & 0.0001 & 1.0999 & 0.0001 & 0.3160 & 0.0000 & 0.7943 & 0.0106 & 1.0487 & 0.0079 & 0.5621 & 0.0063 \\
588.2594 & 0.0061 & 17.0240 & 0.0000 & 3.5265 & 0.0000 & 3.1363 & 0.0000 & 4.1260 & 0.0032 & 1.8779 & 0.0013 & 1.7710 & 0.0005 \\
596.3517 & 0.0000 & 0.3169 & 0.0000 & 0.0013 & 0.0000 & 0.0003 & 0.0000 & 0.5629 & 0.0000 & 0.0365 & 0.0000 & 0.0185 & 0.0000 \\
597.7355 & 0.0000 & 0.4200 & 0.0000 & 0.0013 & 0.0000 & 0.0003 & 0.0000 & 0.6480 & 0.0019 & 0.0364 & 0.0000 & 0.0185 & 0.0000 \\
    \end{tabular}
    \caption{Ideal excitation energies and oscillator energies shown in Fig. \ref{main-fig:noise_low_energy} with the variance and standard deviation of the 10 samples using 10k, 50k, and 100k shots for the respective ideal excitation energies and oscillator strengths.}
    \label{tab:shot_noise}
\end{table}
\end{sidewaystable}

\clearpage